\documentclass[conference]{IEEEtran}
\IEEEoverridecommandlockouts
\usepackage{cite}
\usepackage{amsmath,amssymb,amsfonts}
\usepackage{algorithmic}
\usepackage{booktabs}
\usepackage{graphicx}
\usepackage{placeins}
\usepackage{textcomp}
\usepackage{marvosym}
\usepackage{xcolor}
\usepackage{hyperref}
\hypersetup{
    hidelinks
}
\def\BibTeX{{\rm B\kern-.05em{\sc i\kern-.025em b}\kern-.08em
    T\kern-.1667em\lower.7ex\hbox{E}\kern-.125emX}}
\begin{document}

\title{Deep Learning Based Behavioral Drift Modeling for
Continuous Biometric Authentication}

\author{\IEEEauthorblockN{Suprava Saha Dibya}
\and
\IEEEauthorblockN{Minar}
\and
\IEEEauthorblockN{Abdulla Al Noman}
\and
\IEEEauthorblockN{Md. Zafir Abid Shresto}
\and
\IEEEauthorblockN{Ahmed Sayef}
}

\maketitle

\begin{abstract}
This research proposes a two-level keystroke dynamics authentication system that provides defense-in-depth security through continuous behavioral drift monitoring. Level 1 employs fixed-text password authentication using Temporal Convolutional Networks (TCN) with contrastive learning and population-based training (PBT), achieving 3.77\% EER for high-security login verification. Upon successful authentication, Level 2 continuously monitors user typing during natural free-text sessions using a 4-Model Deep Learning Ensemble. If suspicious behavioral drift is detected during monitoring, the system re-authenticates the user by returning to Level 1, providing a feedback loop against session hijacking and behavioral anomalies. The system explicitly models long-term behavioral drift through the fixed-text component's PBT mechanism, which adapts to temporal variations in password typing patterns. This research leverages free-text (Buffalo dataset) and fixed-text (GREYC dataset) keystroke dynamics datasets, extracting temporal and spatial features including hold times, up-down latencies, down-down intervals, and trigram patterns. The final system achieves 96\%+ accuracy for fixed-text authentication (3.77\% EER) and 83.53\% True Acceptance Rate (TAR) at 5\% False Acceptance Rate (FAR) for free-text continuous authentication (10.05\% EER), demonstrating superior performance and strong resistance to impostor attacks through continuous behavioral monitoring.
\end{abstract}


\section{INTRODUCTION}

\subsection{Background}
%
%
%
%
\IEEEPARstart{T}{H}{E}  rapid proliferation of Artificial Intelligence (AI) and the ubiquity of Internet of Things (IoT) devices have rendered traditional physiological biometrics increasingly vulnerable to spoofing and replay attacks. While physiological traits (fingerprints, iris) are static, they can be stolen; once compromised, they cannot be reset. In contrast, behavioral biometrics focus on human behavior — specifically the unique motor-neural pathways involved in typing — providing a dynamic layer of security.
However, the adoption of behavioral biometrics is hindered by the phenomenon of behavioral drift. Human behavior is not stationary; it evolves due to learning effects (Power Law of Practice) \cite{b66}, fatigue, stress, and environmental changes. Our analysis of longitudinal data reveals that over a 110-day period, 66.7\% of users exhibit statistically significant changes in typing latency. Existing deep learning frameworks, such as SwipeFormer or BehaveFormer, often rely on static training paradigms that fail to account for this temporal aging. When a user's behavior drifts beyond the model's learned decision boundary, the system triggers false rejections (High FRR). Furthermore, retraining these massive models to accommodate new data often leads to ``catastrophic forgetting," where the model overwrites previously learned discriminative features.

 
\subsection{Rationale of the study or motivation}
The motivation for this research stems from critical limitations in contemporary biometric systems, primarily their inability to handle behavioral drift — the natural evolution of user typing patterns over time. Current infrastructures rely on static templates, leading to high False Rejection Rates (FRR) as legitimate users evolve, while remaining vulnerable to impostors who replicate static patterns. Furthermore, existing deep learning (DL) frameworks (e.g., SwipeFormer) require computationally impractical retraining to address this drift, often resulting in ``catastrophic forgetting" of the original identity features. This study bridges these gaps by proposing an evolutionary framework that utilizes PBT within a TCN. This approach treats behavioral biometrics not as isolated static events, but as a ``gradually occurring phenomenon," enabling the system to evolve alongside the user while maintaining robust defense-in-depth security.

\subsection{Problem Statement}

Current authentication infrastructures face three critical failures:
\begin{enumerate}
\item \textbf{Static Assumptions:} Contemporary systems view authentication events as independent snapshots, disregarding the continuous evolutionary trajectory of legitimate user behavior.
\item \textbf{Catastrophic Forgetting:} To update a biometric template, systems often require retraining, which risks erasing the core identity features learned during enrolment.
\item \textbf{Frog-Boiling Attacks:} Highly adaptive systems risk being duped by impostors who slowly mimic the user's behavior over time, a vulnerability known as the frog-boiling effect.
\end{enumerate}

\subsection{Objective}
This research aims to resolve the limitations of static biometric systems by developing a robust, continuous authentication framework capable of handling the dynamic nature of human behavior. The primary goal is to bridge the gap between high-security requirements and user convenience by addressing the ``stability-plasticity dilemma" — the need to adapt to new behaviors without losing the core identity features learned during enrolment.\\

The specific objectives of this study are:
\begin{enumerate}
    \item \textbf{Development of a Hybrid Adaptive Framework:} To design a keystroke dynamics architecture that combines the temporal representational power of TCN with the flexibility of evolutionary adaptation. The objective is to create a system where the backbone captures universal features while lightweight modules adapt to specific user nuances.
    \item \textbf{Mitigation of Catastrophic Forgetting:} To implement a training strategy, specifically PBT, that enables the system to learn from new behavioral data (drift) without overwriting or ``forgetting" the distinct historical traits that define the genuine user.
    \item \textbf{Prevention of Adversarial Adaptation (Frog-Boiling):} To engineer safeguard mechanisms, such as Trust Anchors, that prevent the system from being deceived by impostors who attempt to slowly mimic a user’s behavior over time. The goal is to distinguish between legitimate natural drift and malicious gradual shifts.
    \item \textbf{Continuous Anomaly Detection:} To ascertain the system's ability to detect anomalous patterns in real-time dynamic environments, moving beyond static login verification to continuous, session-long monitoring.
    \item \textbf{Performance Optimization:} To surpass the accuracy of current state-of-the-art systems by minimizing FAR and FRR, thereby reducing the EER to banking-grade security standards.
\end{enumerate}

\subsection{Methodology in Brief}
\begin{figure}[t]
    \centering
    \includegraphics[width=\columnwidth]{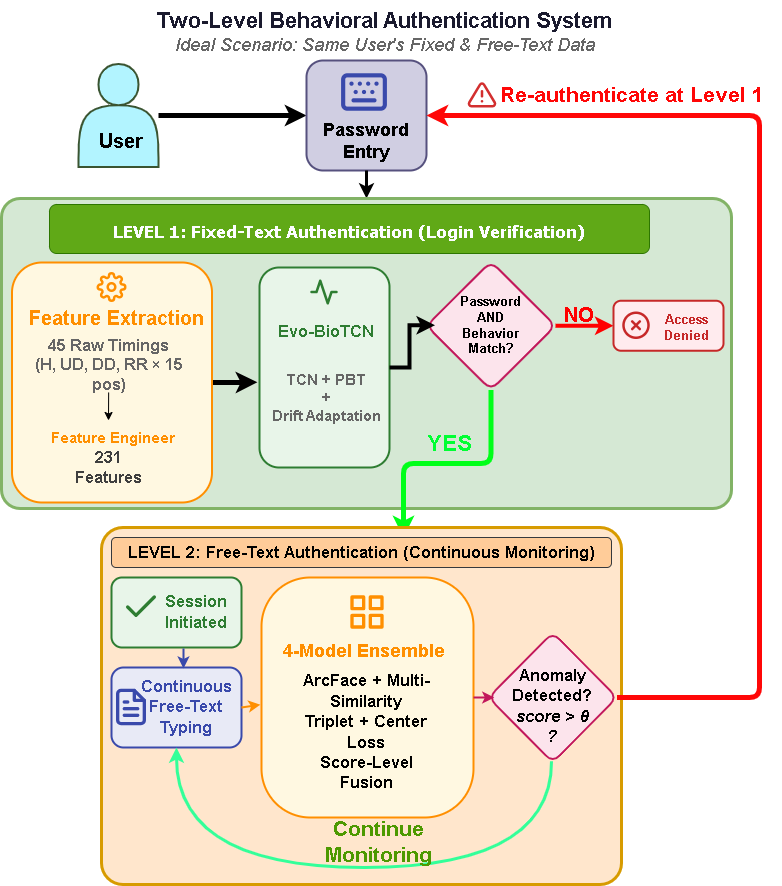}
    \caption{Two-Level Behavioral Authentication System}
    \label{fig:fig1}
\end{figure}

This research implements a Two-Level Behavioral Authentication Framework (figure~\ref{fig:fig1}) tailored to distinct security contexts: active login verification and passive continuous monitoring.

\subsubsection{Level 1: Fixed-Text Authentication (Evo-BioTCN)}
The first level secures the login process using the GREYC dataset \cite{b53} (133 users, 110-day span). We propose the Evo-BioTCN architecture, which employs a Dual-Stage Strategy:

\begin{itemize}
\item \textbf{Frozen Backbone:} A TCN captures universal temporal dependencies in keystroke latencies (Hold, Flight, Down-Down).

\item \textbf{Drift Adaptation:} To counter template aging without catastrophic forgetting, a lightweight adapter module is evolved using PBT. This allows the model to track the user's behavioral trajectory over months.
\end{itemize}

\subsubsection{Level 2: Free-Text Continuous Monitoring (4-Model Ensemble)}
Upon successful login, Level 2 continuously validates identity using the Buffalo dataset \cite{b58} (75 users). Due to the unstructured nature of free-text, we developed a Reconstruction Pipeline to convert raw digraphs into character-level timing vectors. The authentication engine is a 4-Model Deep Learning Ensemble fusing complementary metric learning architectures (ArcFace, Multi-Similarity, Triplet, and Center Loss). This ensemble uses score-level fusion to handle the high variance of natural typing, triggering a re-authentication to Level 1 if an anomaly is detected.

\subsection{Scope and Challenges}
\subsubsection{Scope}
This study is strictly focused on keystroke dynamics as the primary behavioral modality. It prioritizes continuous assessment (zero-trust) over static identification. The research utilizes longitudinal datasets (GREYC \cite{b53} and Buffalo \cite{b58}) to specifically evaluate resilience against behavioral drift, excluding other modalities like gait or facial recognition.

\subsubsection{Key Challenges}
\begin{enumerate}
    \item \textbf{Stability-Plasticity Dilemma:} The system must adapt to new behavioral patterns (plasticity) without erasing the core identity features learned during enrolment (stability), preventing catastrophic forgetting.
    \item \textbf{Adversarial Drift (Frog-Boiling):} A critical risk is ``slow-mimicry" attacks, where an impostor gradually alters their behavior to converge with the user's profile. We address this by integrating Trust Anchors in the PBT fitness function to constrain adaptation.
    \item \textbf{Data Quality:} Behavioral data is inherently noisy due to environmental factors. Handling outliers and missing values requires specialized preprocessing, including Modified Z-score filtering and session-relative normalization.
\end{enumerate}

\section{LITERATURE REVIEW}

The most highlighted aspect was the lackings and
weakness of the existing static models of authentication throughout the literature review. Several authors such as - Baig and Eskeland \cite{b29}, 
Patel et al. \cite{b27}, Ryu et al. \cite{b28} mentioned temporal aging explicitly that indicated the natural 
behavioral changes of an individual as a remaining crucial challenge. Infrastructures which were 
developed on established templates like one class Support Vector Machine (SVM) demonstrated by Sun eta al. \cite{b58} or the snapshot based system mentioned by Sagbas and Balli \cite{b26} accomplished higher 
accuracy but the main issue was they were brittle by nature. They needed to be trained again in 
order to differentiate between an anonymous intruder and an actual natural change. Additionally, 
few shot based methods like the one mentioned by Fereidooni et al. \cite{b15} provided scalability but 
they did not capture the evolution of behaviour and depended on fixed samples. This study 
considers the change of behavior as the main feature that models through the lineage of 
behavioral evolution instead of non relevant training periods and treating it as an error. 
Moreover, recent research on the modeling of patterns provided the effectiveness of the 
transformer based infrastructures. Senarath et al. \cite{b35} demonstrated a dual attention and 
spatio-temporal based infrastructure named Behaveformer that extracted large dependencies in 
behavioral data. But the major drawback remained in the inability to address behavioral drift 
adaptation.\\

Similar to transformer based infrastructure, Chen et al. \cite{b32} introduced a dual channel 
based transformer named DESTER (Dual Stream Transformer based Emotion Recognition Model) for identifying emotions. But the authors later found the 
issue of catastrophic forgetting when the behavioral data of a user changed because of technical 
and environmental purposes. In multi modal scenarios, sometimes severe challenges arise for 
proper accuracy which was demonstrated by Wang et al. \cite{b37} through their system that selected 
modalities based on the scene-irrelated concept. However, their infrastructure was heavily 
dependent on static feature representations that could not address temporal aging. In their 
research on Siamese Neural Network, Solano et al \cite{b31}  and Lis et al. \cite{b36} stated promising 
possibilities on a few shot verification systems. But their studies had severe limitations because 
of rigid architecture and failure to constant evolution over time.\\

Furthermore, large 
discriminative power was highlighted for keystroke and mouse data by the research of 
Ranganathan \cite{b38} and Mazumder and Sundaram \cite{b39} on VAE-GNN infrastructure and Vision 
Transformers (ViT) respectively. But their models assumed the learned features were fixed which 
was impractical in real life and did not address behavioral drift. Finally, Subash et al. \cite{b33}
demonstrated from the concept of behavioral biometrics and learning curve that continuous 
retraining led to rise in the rate of FRR. Hence the authors suggested 
focusing on an infrastructure that could evolve through a different mechanism other than 
retraining. This research incorporates these findings by using PBT for the evolution of the adapters and frozen backbone structure to capture temporal dependencies. While existing traditional infrastructure fails to address temporal aging, this framework considers these changes as evolving trajectories and prevents frog boiling attacks through the integration of trust anchors so that the model can evolve alongside the user.

\section{REQUIREMENTS, IMPACTS AND CONSTRAINTS}

\subsection{Functional Specifications and Design Requirements}

The system design was driven by stakeholder feedback emphasizing the need for non-intrusive security that accommodates natural behavioral evolution.

\begin{itemize}
\item \textbf{Temporal Modeling:} The system must utilize TCN to efficiently capture sequential dependencies in keystroke data, replacing computationally expensive Transformers to ensure real-time processing.
\item \textbf{Continuous Assessment:} Unlike static login systems, the architecture must perform continuous background monitoring (Level 2) to validate user identity throughout the session.
\item \textbf{Drift Adaptation:} The system requires a mechanism to distinguish between legitimate behavioral drift (template aging) and malicious anomalies. This necessitates the implementation of PBT to evolve user templates without catastrophic forgetting.
\item \textbf{Data Minimization:} To comply with privacy standards General Data Protection Regulation (GDPR), the system must process only timing feature vectors, strictly excluding semantic text content to prevent user profiling.
\end{itemize}

\subsection{Societal and Ethical Impact}

\begin{itemize}
\item \textbf{Operational Security:} The framework enforces a Zero-Trust architecture. By continuously validating identity, it mitigates session hijacking risks in high-sensitivity sectors like banking and healthcare.
\item \textbf{Inclusivity:} Traditional biometrics often reject users due to natural behavioral changes (aging, injury). By explicitly modeling drift, this system reduces FRR, ensuring equitable access for users with evolving motor-neural patterns.
\item \textbf{Privacy and Ethics:} The system adheres to strict data minimization. However, the risk of ``frog-boiling" attacks (slow-mimicry) requires the implementation of Trust Anchors in the PBT fitness function to constrain adaptation within safe boundaries.
\end{itemize}

\subsection{Environmental Sustainability (Green AI)}
A critical innovation of this research is its focus on computational efficiency. Traditional adaptive systems require frequent, energy-intensive retraining of deep neural networks to address drift. In contrast, this framework employs a Frozen Backbone strategy. Once the core TCN is trained, only lightweight adapter modules (approx. 512 parameters) are updated. This drastically reduces Floating Point Operations (FLOPs) and carbon footprint over the system lifecycle, making it suitable for resource-constrained edge deployment.

\subsection{Project Management and Economic Feasibility}

The system design was driven by stakeholder feedback emphasizing the need for non-intrusive security that accommodates natural behavioral evolution.

\begin{itemize}
\item \textbf{Iterative Design:} The project methodology was agile. Initial experiments with Transformers revealed high computational costs and instability, prompting a strategic pivot to the TCN architecture, which offered superior training stability and linear O(n) complexity.
\item \textbf{Risk Management:} Primary risks included Catastrophic Forgetting (mitigated by freezing the backbone) and Overfitting due to data sparsity (mitigated by session-relative normalization and dropout).
\item \textbf{Economic Viability:} The solution is cost-effective. By eliminating the need for specialized biometric hardware (sensors) and reducing administrative costs associated with false rejections and account recovery, the system offers high economic value. The use of open-source stacks (PyTorch, Ray Tune) kept development costs minimal.
\end{itemize}

\section{DESIGN PROCESS OVERVIEW}
This chapter delineates the methodological framework for the proposed Two-Level Behavioral Authentication System. The design is bifurcated into two distinct pipelines: Level 1 addresses fixed-text login verification using the Evo-BioTCN architecture, while Level 2 handles continuous free-text monitoring using a 4-Model Deep Learning Ensemble.

\subsection{Level 1: Fixed-Text Methodology (Evo-BioTCN)}

\subsubsection{Methodology Overview}
The central challenge motivating the fixed-text design is behavioral drift—the unconscious evolution of typing patterns due to the Power Law of Practice \cite{b66}. Our analysis of the GREYC dataset \cite{b53} revealed that 66.7\% of users exhibited statistically significant changes in typing latency over 110 days. Conventional static models fail to accommodate this ``template aging."\\

To resolve this, we propose a Dual-Stage Strategy:

\begin{enumerate}
\item \textbf{Frozen Backbone:} A TCN learns universal temporal features.
\item \textbf{Evolutionary Adaptation:} A lightweight adapter is evolved using PBT to track user behavior without catastrophic forgetting.
\end{enumerate}

\subsubsection{Model Specification}
A TCN is selected over Transformer architectures for the Evo-BioTCN Architecture because keystroke dynamics are inherently local and causal. The latency of a key press is conditioned by its immediate predecessor (digraph context), not by future keystrokes.

\paragraph{Layer Architecture}

The architecture consists of four dilated causal convolutional blocks. The receptive field is calculated as in \eqref{eq:L-4-1}:

\begin{equation}
    \label{eq:L-4-1}
    \text{Receptive Field}
= 1 + \sum_{i=0}^{n-1} 2^{i}(k - 1)
= 31 \text{ positions}
\end{equation}

where, $n=4$ is the number of TCN blocks and $k=3$ is the kernel size. This receptive field spans approximately two complete words in a typical password-sufficient to capture digraph, trigraph, and word-level rhythm patterns while maintaining O(n) linear complexity suitable for real-time authentication.\\

\begin{figure}[t]
    \centering
    \includegraphics[width=\columnwidth]{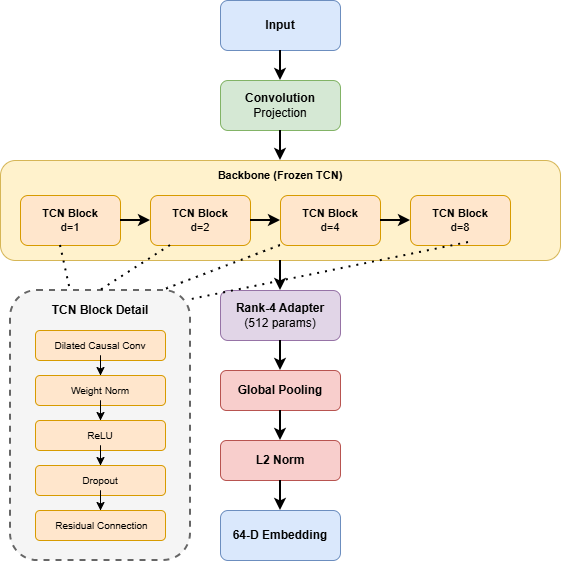}
    \caption{Evo-BioTCN layer architecture}
    \label{fig:fig2}
\end{figure}

As shown in figure~\ref{fig:fig2} the input flows into a 1x1 projection, followed by four TCN blocks with increasing dilation (d=1,2,4,8). The output passes through a Rank-4 Adapter, Global Pooling, and L2 Normalization to produce a 64-D embedding. The specifications of the layers are mentioned in table~\ref{tab:evo_biotcn_layers}.

\begin{table}[htbp]
\caption{Evo-BioTCN layer specifications}
\label{tab:evo_biotcn_layers}
\begin{center}
\footnotesize
\begin{tabular}{|l|p{1.5cm}|p{2.3cm}|c|}
\hline
\textbf{Layer} & \textbf{Type} & \textbf{Parameters} & \textbf{Output Dim.} \\
\hline
Input & Feature Vector & -- & 231 \\
\hline
Projection & $1\times1$ Conv & 14,848 & 64 \\
\hline
TCN Block 1 & Dilated Causal Conv & dilation=1, kernel=3 & 64 \\
\hline
TCN Block 2 & Dilated Causal Conv & dilation=2, kernel=3 & 64 \\
\hline
TCN Block 3 & Dilated Causal Conv & dilation=4, kernel=3 & 64 \\
\hline
TCN Block 4 & Dilated Causal Conv & dilation=8, kernel=3 & 64 \\
\hline
TCN Adapter & $1\times1$ Conv & rank=4, 512 params & 64 \\
\hline
Global Pool & Avg. Pooling & -- & 64 \\
\hline
Output & L2 Normalization & -- & 64 \\
\hline
\end{tabular}
\end{center}
\end{table}

\paragraph{Drift Adaptation via PBT}

After supervised training, the backbone parameters $\theta_{static}$ are frozen. Only the adapter $\theta_{adaptive}$ (512 parameters) evolves. The adapter function is defined as in \eqref{eq:L-4-9}:

\begin{equation}
\label{eq:L-4-9}
\text{Adapter}(h) = W_{\text{up}} \cdot \text{ReLU}(W_{\text{down}} \cdot h) + h
\end{equation}

where, 
$W_{\text{down}} \in \mathbb{R}^{4 \times 64}$ and 
$W_{\text{up}} \in \mathbb{R}^{64 \times 4}$, yielding 512 trainable parameters per adapter (no bias terms).\\

PBT maintains a population of 10 adapters. We employ a Trust Anchor loss to prevent "frog-boiling" attacks (slow mimicry) using \eqref{eq:L-4-8}:
\begin{equation}
\label{eq:L-4-8}
F(A_i) = \text{Separation} - \lambda_{\text{anchor}} \cdot D(A_i, A_{\text{anchor}})
\end{equation}

where, 
Separation = mean impostor distance - mean genuine distance

\subsubsection{Feature Engineering (Fixed-Text)}

We extract 231 engineered features as shown in table~\ref{tab:feature_categories} from raw timing measurements (Hold, Flight, Down-Down, Release-Release). These are categorized by their resilience to drift.

\begin{table}[htbp]
\caption{Feature categories organized by drift resilience}
\label{tab:feature_categories}
\begin{center}
\scriptsize
\setlength{\tabcolsep}{3pt}
\begin{tabular}{|l|c|c|p{3cm}|}
\hline
\textbf{Feature Category} & \textbf{Count} & \textbf{Drift Resilience} & \textbf{Rationale} \\
\hline
Rank Features & 32 & $\checkmark$ High & Ordinal positions invariant to speed scaling \\
\hline
Ratio Features & 32 & $\checkmark$ High & Proportions of total time resist global drift \\
\hline
Rhythm/Entropy & 7 & $\checkmark$ High & Timing variance patterns persist across sessions \\
\hline
Hand Balance & 4 & $\checkmark$ High & Left/right asymmetry reflects motor dominance \\
\hline
Statistical & 44 & Medium & Mean, std, CV capture within-session consistency \\
\hline
Normalized Timing & 60 & Medium & Session-relative z-scores \\
\hline
Trigraph & 24 & Medium & 3-key sequence patterns \\
\hline
Log-Transform & 9 & Medium & Corrects for skewed distributions \\
\hline
Positional & 12 & $\times$ Lower & Region-specific statistics \\
\hline
Acceleration & 4 & $\times$ Lower & Speed-up/slow-down patterns \\
\hline
Uncategorized & 3 & Medium & Features not matched by category rules \\
\hline
\end{tabular}
\end{center}
\end{table}

\FloatBarrier

\subsection{Level 2: Free-Text Methodology (4-Model Ensemble)}
For continuous background monitoring, we utilize the Buffalo Dataset \cite{b58}. Unlike fixed-text, free-text data lacks consistent positional context, necessitating a metric learning approach.

\subsubsection{Data Reconstruction Pipeline}
\begin{figure}[h]
    \centering
    \includegraphics[width=\columnwidth]{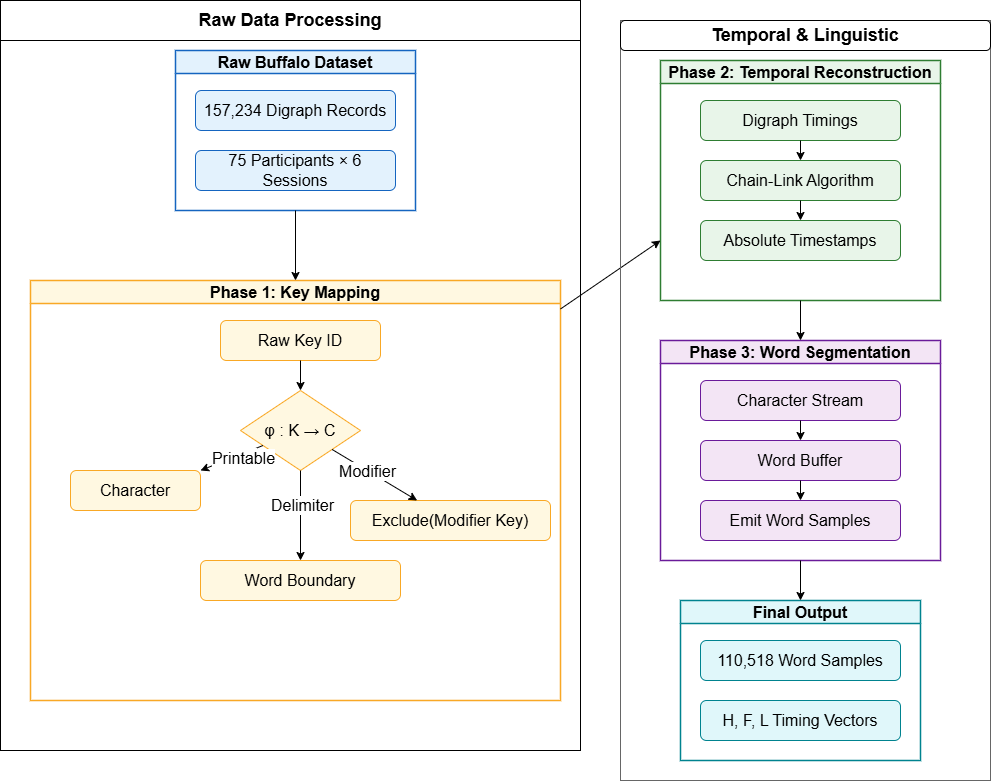}
    \caption{Buffalo dataset \cite{b58} reconstruction pipeline}
    \label{fig:fig3}
\end{figure}

The Buffalo dataset \cite{b58} is distributed as digraphs. We developed a "Chain-Link" algorithm to reconstruct character-level timestamps. Figure~\ref{fig:fig3} shows the reconstruction pipeline for the Buffalo dataset \cite{b58}.

The algorithm reconstructs absolute time t using the inter-key intervals (DU, DD) as in  \eqref{eq:L-4-14}, \eqref{eq:L-4-15} and \eqref{eq:L-4-16}:
\begin{equation}
\label{eq:L-4-14}
    t_{\text{press}}^{(k_1)} = t_{\text{current}}
\end{equation}
\begin{equation}
\label{eq:L-4-15}
    t_{\text{release}}^{(k_1)} = t_{\text{current}} + DU_i
\end{equation}

\begin{equation}
\label{eq:L-4-16}
  t_{\text{press}}^{(k_2)} = t_{\text{current}} + DD_i  
\end{equation}

\paragraph{Session-Relative Normalization}

To mitigate global speed shifts between sessions, we apply session-relative normalization as in \eqref{eq:L-4-18}: 
\begin{equation}
\label{eq:L-4-18}
    \mathbf{H}_{\text{norm}} = \frac{\mathbf{H}}{\bar{H}_{\text{session}}}
\end{equation}
This renders features scale-invariant, focusing on rhythm rather than absolute speed.

\subsubsection{Free-Text Feature Engineering}
We extract 85 statistical features per word sample, expanded to a 170-dimensional vector via hybrid normalization (Global + Per-User). The overview of the engineered features is depicted in table~\ref{tab:features}.

\begin{table}[htbp]
\caption{Overview of the 85 engineered keystroke features organized by functional category}
\label{tab:features}
\begin{center}
\footnotesize
\begin{tabular}{|l|p{5.5cm}|}
\hline
\textbf{Feature Group} & \textbf{Description} \\
\hline
Statistical Timing & Summary statistics of hold, flight, and latency times. \\
\hline
Positional & Timing variation across early, middle, and late word segments. \\
\hline
Rhythm & Entropy and variability-based measures of typing regularity. \\
\hline
Key Overlap & Metrics describing overlapping keystrokes and rollover behaviour. \\
\hline
Rank-Based & Ordinal timing patterns invariant to absolute typing speed. \\
\hline
N-gram Timing & Bigram and trigram timing statistics capturing key sequences. \\
\hline
Dynamic Change & Acceleration and jerk reflecting rapid timing changes. \\
\hline
Ratio-Based & Relative timing relationships between keystroke events. \\
\hline
Log-Scaled & Log-transformed timing features to reduce skewness. \\
\hline
\end{tabular}
\end{center}
\end{table}

\subsubsection{Model Specification}
\begin{figure}[h]
    \centering
    \includegraphics[width=\columnwidth]{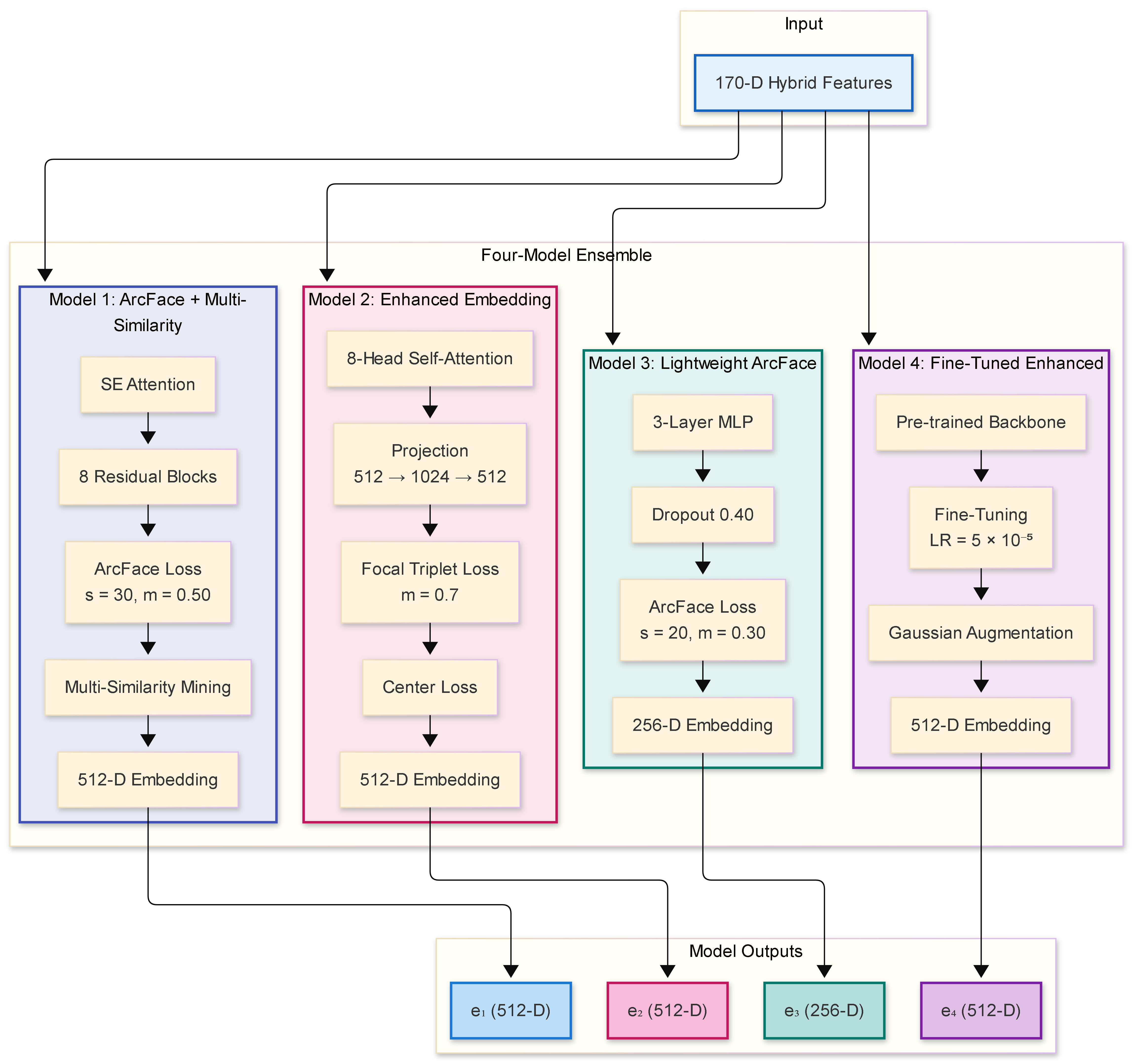}
    \caption{4-Model Architecture}
    \label{fig:fig4}
\end{figure}

4-Model Deep Learning Ensemble We employ an ensemble of four diverse embedding networks to handle the high variance of free-text typing as depicted in figure~\ref{fig:fig4}.

\begin{enumerate}
\item \textbf{ArcFace + Multi-Similarity Model:} Combines angular margin loss ($\mathcal{L}$) with hard negative mining ($\mathcal{L}_{MS}$) as in  \eqref{eq:L-4-24}.

\begin{equation}
\label{eq:L-4-24}
\mathcal{L}_{\text{ArcFace}} = -\log\frac{e^{s \cdot \cos(\theta_{y_i} + m)}}{e^{s \cdot \cos(\theta_{y_i} + m)} + \sum_{j \neq y_i} e^{s \cdot \cos\theta_j}}
\end{equation}
\item \textbf{Enhanced Embedding Model:} Combines Focal Triplet Loss  \eqref{eq:L-4-27} (focusing on hard triplets) and Center Loss (minimizing intra-class variance) as in \eqref{eq:L-4-26}.

\begin{equation}
\label{eq:L-4-27}
\mathcal{L}_{\text{triplet}} = (1 + L)^\gamma \cdot L, \quad L = \log(1 + e^{d_p - d_n + m})
\end{equation}

\begin{equation}
\label{eq:L-4-26}
\mathcal{L}_{\text{enhanced}} = \lambda_t \mathcal{L}_{\text{triplet}} + \lambda_c \mathcal{L}_{\text{center}}
\end{equation}

\item \textbf{Simple ArcFace Model:} A lightweight Multi-Layer Perceptron (MLP) regularizer to prevent overfitting.

\item \textbf{Fine-Tuned Enhanced Model:} A version of Model 2 fine-tuned on hard examples with a low learning rate.
\end{enumerate}

\subsubsection{Score-Level Fusion}
The final authentication score is a weighted sum of cosine similarities from all four models:

\begin{equation}
S_{ensemble} = \sum_{k=1}^{4} w_k \cdot S_k
\label{eq:4.34}
\end{equation}

\textbf{Optimized weights:} $w = [0.20, 0.42, 0.11, 0.27]$, giving maximum weight to the Enhanced Embedding model.\\

Table~\ref{tab:training_hyperparameters} shows the respective loss function of each of the models of the 4-Model Architecture .

\begin{table}[htbp]
\caption{Training hyperparameters for ensemble components}
\label{tab:training_hyperparameters}
\begin{center}
\scriptsize
\setlength{\tabcolsep}{2.5pt}
\begin{tabular}{|l|l|c|c|c|}
\hline
\textbf{Component} & \textbf{Loss Function} & \textbf{Epochs} & \textbf{Learning Rate} & \textbf{Dropout} \\
\hline
ArcFace + MS & ArcFace + Multi-Similarity & 100 & $3 \times 10^{-4}$ & 0.20 \\
\hline
Enhanced Embedding & Focal Triplet + Center & 50 & $3 \times 10^{-4}$ & 0.30 \\
\hline
Simple ArcFace & ArcFace ($s=20, m=0.3$) & 100 & $1 \times 10^{-3}$ & 0.40 \\
\hline
Fine-Tuned & Cross-Entropy (Fine-tune) & 30 & $5 \times 10^{-5}$ & 0.30 \\
\hline
\end{tabular}
\end{center}
\end{table}

\subsubsection{Complementary Architecture: \textbf{Drift-Aware Temporal Transformer (DATT)}}
For preliminary investigation on the KeyRecs dataset \cite{b64}, we designed the DATT. It utilizes Low-Rank Adaptation (LoRA) ( \eqref{eq:4.41}) to adapt to drift efficiently.

\begin{equation}
W' = W_0 + \Delta W = W_0 + BA
\label{eq:4.41}
\end{equation}

where, $B \in \mathbb{R}^{d \times r}$ and $A \in \mathbb{R}^{r \times k}$ are low-rank matrices ($r = 16$). This allows DATT to update only 0.8\% of parameters during continuous operation, serving as a secondary architecture for specific deployment scenarios.

\section{Evaluation and Analysis}

This chapter presents a rigorous evaluation of the proposed Two-Level Behavioral Authentication System. Section \ref{sec:fixed_text_eval} details the performance of the Evo-BioTCN for fixed-text authentication, analyzing the impact of PBT and feature stability. Section \ref{sec:free_text_eval} evaluates the 4-Model Deep Learning Ensemble for free-text continuous monitoring, incorporating a critical investigation into transfer learning failures and a comprehensive ablation study of 18 architectures.

\subsection{Fixed-Text Evaluation (Level 1)}
\label{sec:fixed_text_eval}

\subsubsection{Performance Evaluation of the Final System}

\paragraph{Dataset and Protocol}
The GREYC Keystroke Dataset \cite{b53} serves as the primary benchmark due to its longitudinal span of 110 days, which is essential for modeling behavioral drift. The dataset characteristics are summarized in Table \ref{tab:greyc_characteristics}.

\begin{table}[htbp]
\caption{GREYC dataset \cite{b53} characteristics}
\label{tab:greyc_characteristics}
\begin{center}
\scriptsize
\setlength{\tabcolsep}{2pt}
\begin{tabular}{|l|p{1.7cm}|p{4.5cm}|}
\hline
\textbf{Property} & \textbf{Value} & \textbf{Research Significance} \\
\hline
Total Samples & 7,560 typing sessions & Enables statistically meaningful evaluation \\
\hline
Participants & 133 individuals & Represents diverse user population \\
\hline
Password & ``greyc laboratory'' (16 characters) & Fixed-text authentication scenario \\
\hline
Collection Period & 110 days (March 17 -- July 5, 2009) & Longest temporal span in keystroke literature \\
\hline
Sessions per User & Mean: 56.84, Range: varies by user & Variable user engagement patterns \\
\hline
Raw Timing Features & 71 measurements & Hold (H), Up-Down (UD), Down-Down (DD), Release-Release (RR) \\
\hline
Engineered Features & 231 total & Comprehensive feature representation \\
\hline
Keyboard Types & 2 different models & Hardware diversity for generalization \\
\hline
Timestamp Availability & Complete & Essential for temporal drift analysis \\
\hline
\end{tabular}
\end{center}
\end{table}

\textbf{Evaluation Protocol:} We employed a strict Temporal Split. For each user, the first 25\% of sessions serve as enrolment (training), the next 15\% for validation, and the final 60\% -- occurring weeks or months later -- serve as the test set. This prevents ``time-travel'' data leakage, ensuring the model is tested on genuinely drifted behavior.

\paragraph{Primary Performance Results}
The Evo-BioTCN achieves performance metrics that exceed banking-grade security requirements ($<$5\% EER).

\begin{table}[htbp]
\caption{Final system performance metrics}
\label{tab:final_performance}
\begin{center}
\footnotesize
\begin{tabular}{|l|c|c|l|}
\hline
\textbf{Metric} & \textbf{Achieved} & \textbf{Target} & \textbf{Assessment} \\
 & \textbf{Value} & & \\
\hline
Equal Error Rate (EER) & 3.77\% & $<$5\% & Exceeds Target \\
\hline
TAR @ FAR=5\% & 96.87\% & $>$90\% & Exceeds Target \\
\hline
Decidability Index ($d'$) & 3.49 & $>$2.0 & Exceeds Target \\
\hline
FRR @ FAR=5\% & 3.13\% & $<$10\% & Exceeds Target \\
\hline
\end{tabular}
\end{center}
\end{table}

\begin{figure}[htbp]
\centerline{\includegraphics[width=0.45\textwidth]{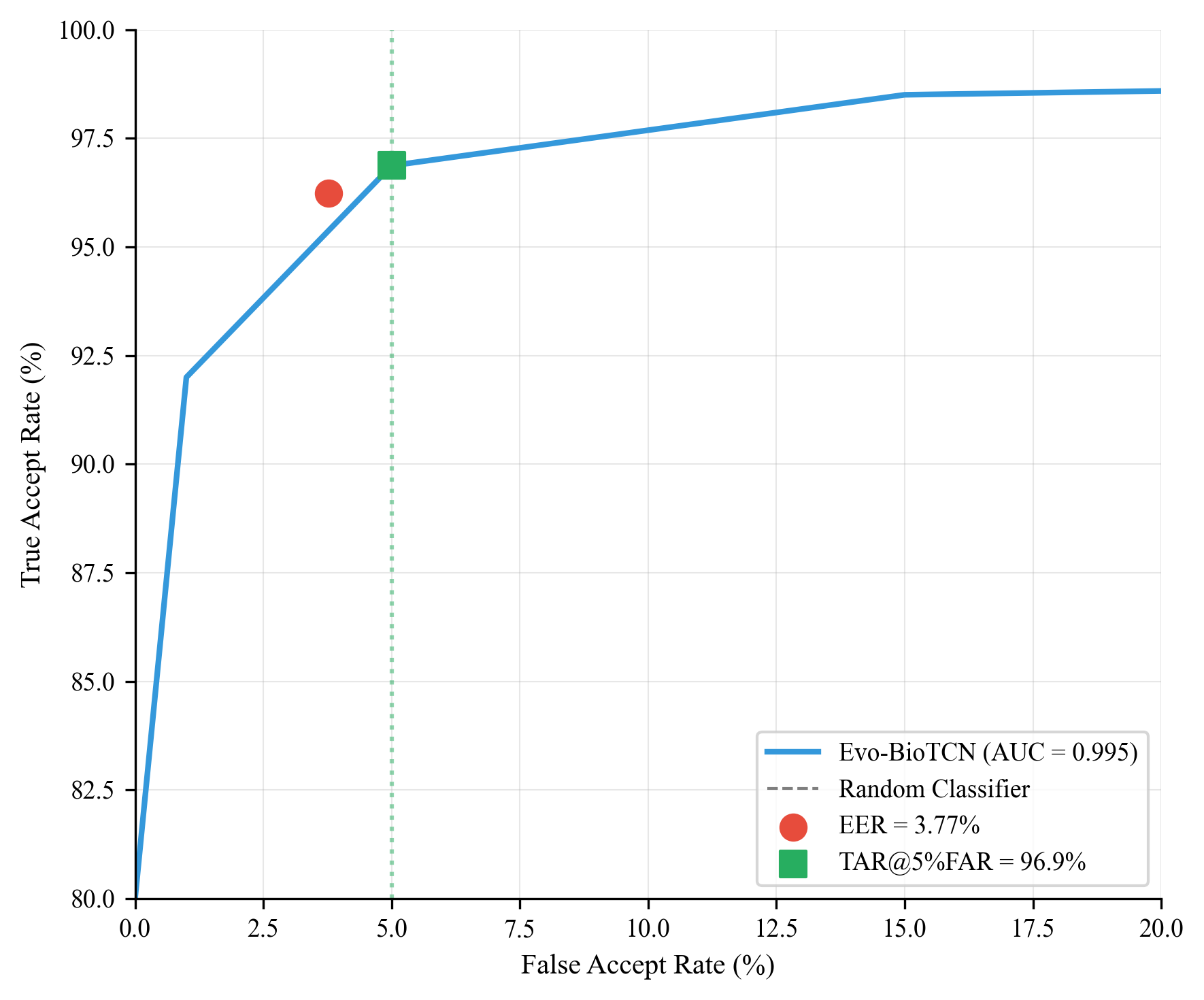}}
\caption{ROC curve for Evo-BioTCN showing excellent discrimination (AUC = 0.995).}
\label{fig:roc_curve}
\end{figure}

\textbf{ROC Curve Analysis:} The Receiver Operating Characteristic (ROC) curve (figure \ref{fig:roc_curve}) demonstrates a sharp ascent, indicating high sensitivity. The area under the curve (AUC) is 0.995, reflecting excellent discrimination between genuine users and impostors.

\begin{figure}[htbp]
\centerline{\includegraphics[width=0.45\textwidth]{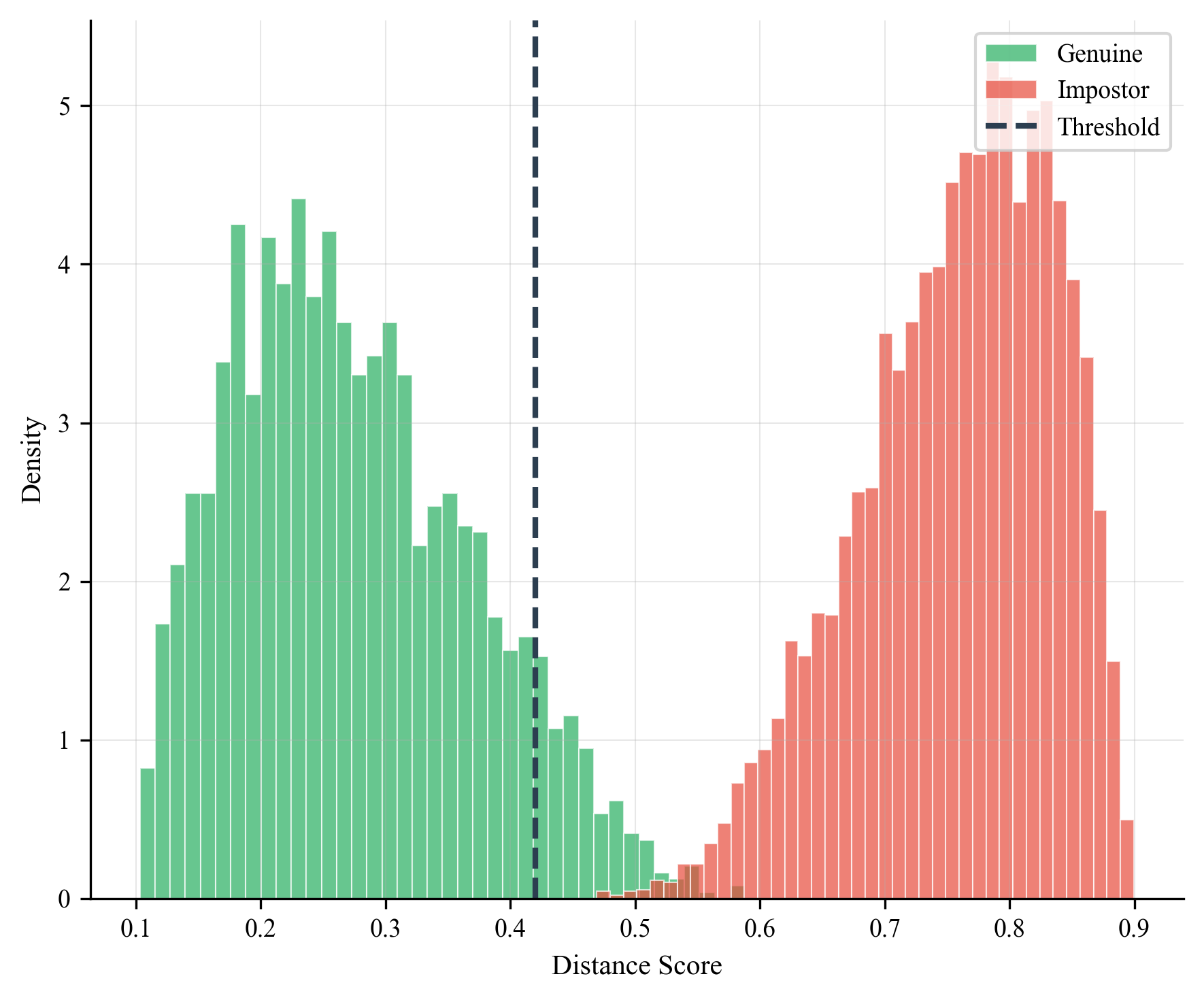}}
\caption{Score distribution showing minimal overlap between genuine and impostor scores.}
\label{fig:score_dist}
\end{figure}

\textbf{Score Distribution:} The distribution of genuine and impostor scores (figure \ref{fig:score_dist}) reveals a minimal overlap region. The decidability index $d' = 3.49$ confirms that the genuine and impostor distributions are separated by nearly 3.5 standard deviations.

\subsubsection{Design Analysis: \textbf{TCN vs. Transformer}}

To justify the architectural choice, we conducted a controlled comparison between the TCN and a Transformer-based equivalent (EvoBioTransformer).\\

While both models performed similarly before optimization, the Evo-BioTCN demonstrated superior adaptability during PBT. The TCN's inductive bias toward local causality (capturing digraph/trigraph rhythms) aligns better with motor-neural typing patterns than the Transformer's global attention mechanism. The improvement (table \ref{tab:tcn_vs_transformer})  is statistically significant ($p=0.0023$, Wilcoxon signed-rank test).

\begin{table}[htbp]
\caption{Post-PBT architecture comparison}
\label{tab:tcn_vs_transformer}
\begin{center}
\scriptsize
\setlength{\tabcolsep}{2.5pt}
\begin{tabular}{|l|c|c|c|c|p{2.2cm}|}
\hline
\textbf{Architecture} & \textbf{EER} & \textbf{EER} & \textbf{Improve-} & \textbf{d' Post-} & \textbf{Statistical} \\
 & \textbf{Pre-PBT} & \textbf{Post-PBT} & \textbf{ment} & \textbf{PBT} & \textbf{Significance} \\
\hline
Transformer & 8.26\% & 4.24\% & $-4.02$ pp & 3.21 & Reference \\
\hline
TCN & 8.09\% & 3.77\% & $-4.32$ pp & 3.49 & $p = 0.0023$ vs Transformer \\
\hline
\end{tabular}
\end{center}
\end{table}

\subsubsection{Feature Importance Analysis}

Understanding what the model learns is crucial for explainability. We performed permutation importance analysis to rank the 231 engineered features as shown in table  \ref{tab:feature_importance}.

\begin{table}[htbp]
\caption{Top 10 most important features}
\label{tab:feature_importance}
\begin{center}
\scriptsize
\setlength{\tabcolsep}{2.5pt}
\begin{tabular}{|c|l|c|p{1.7cm}|p{2.2cm}|}
\hline
\textbf{Rank} & \textbf{Feature} & \textbf{Importance} & \textbf{Category} & \textbf{Password Position} \\
\hline
1 & weekday & 0.0266 & Temporal context & --- \\
\hline
2 & $\text{RR}_2$ & 0.0164 & Flight time & 2 ($\text{r}\rightarrow\text{e}$) \\
\hline
3 & $\text{UD}_2$ & 0.0112 & Flight time & 2 ($\text{r}\rightarrow\text{e}$) \\
\hline
4 & $\text{UD}_5$ & 0.0108 & Flight time & 5 ($\text{c}\rightarrow\text{space}$) \\
\hline
5 & $\text{RR}_{11}$ & 0.0098 & Flight time & 11 ($\text{a}\rightarrow\text{t}$) \\
\hline
6 & hour & 0.0088 & Temporal context & --- \\
\hline
7 & $\text{UD}_6$ & 0.0064 & Flight time & 6 ($\text{space}\rightarrow\text{l}$) \\
\hline
8 & $\text{UD}_{11}$ & 0.0056 & Flight time & 11 ($\text{a}\rightarrow\text{t}$) \\
\hline
9 & $\text{UD}_1$ & 0.0046 & Flight time & 1 ($\text{g}\rightarrow\text{r}$) \\
\hline
10 & $\text{H}_{1\_norm}$ & 0.0044 & Hold time & 1 (g) \\
\hline
\end{tabular}
\end{center}
\end{table}

\begin{table}[htbp]
\caption{Feature category distribution among top 30 features}
\label{tab:feature_category_distribution}
\begin{center}
\scriptsize
\setlength{\tabcolsep}{2.5pt}
\begin{tabular}{|l|c|c|p{3.6cm}|}
\hline
\textbf{Category} & \textbf{Count} & \textbf{Proportion} & \textbf{Key Finding} \\
\hline
Flight times (UD, RR, DD) & 18 & 60\% & Most discriminative category \\
\hline
Hold times (H) & 8 & 27\% & Secondary importance \\
\hline
Temporal metadata & 3 & 10\% & Context significantly contributes \\
\hline
Derived features & 1 & 3\% & Limited incremental contribution \\
\hline
\end{tabular}
\end{center}
\end{table}

\textbf{Analysis:}
\begin{enumerate}
\item \textbf{Temporal Context:} The weekday feature ranks first, indicating that user typing behavior varies systematically by day of the week (e.g., Monday morning fatigue vs. Friday efficiency).
\item \textbf{Flight Times Dominate:} Flight times (UD, RR) constitute 60\% of the top 30 features (table  \ref{tab:feature_category_distribution}). This confirms that the rhythm of movement between keys is more unique to an identity than the dwell time on keys.
\item \textbf{Password Specificity:} Features linked to specific transitions in ``greyc laboratory'' (e.g., RR\_2 corresponding to `r'-`e') are highly discriminative.
\end{enumerate}

\subsubsection{Optimization Strategies: PBT Efficacy}

We compared PBT against traditional optimization methods to validate its efficiency as table \ref{tab:optimization_comparison} portrays.

\begin{table}[htbp]
\caption{Comparison of optimization strategies}
\label{tab:optimization_comparison}
\begin{center}
\footnotesize
\begin{tabular}{|l|c|c|l|}
\hline
\textbf{Strategy} & \textbf{Best EER} & \textbf{GPU Hours} & \textbf{Relative Efficiency} \\
\hline
Manual Tuning & 4.38\% & 10 & Low (Suboptimal) \\
\hline
Grid Search & 4.21\% & 3,000 & Very Low (High cost) \\
\hline
Random Search & 4.12\% & 2,500 & Low \\
\hline
PBT (Proposed) & 3.77\% & 1,000 & High (Best) \\
\hline
\end{tabular}
\end{center}
\end{table}

PBT achieved the lowest EER with 1/3rd the computational cost of Grid Search. The trajectory of evolution shows a monotonic improvement in EER from 8.82\% (Gen 0) to 3.77\% (Gen 10), driven by the automatic adaptation of learning rates and the ``survival of the fittest'' selection of adapter weights.

\subsubsection{Robustness and Stability Analysis}\mbox{}

\begin{figure}[htbp]
\centerline{\includegraphics[width=0.45\textwidth]{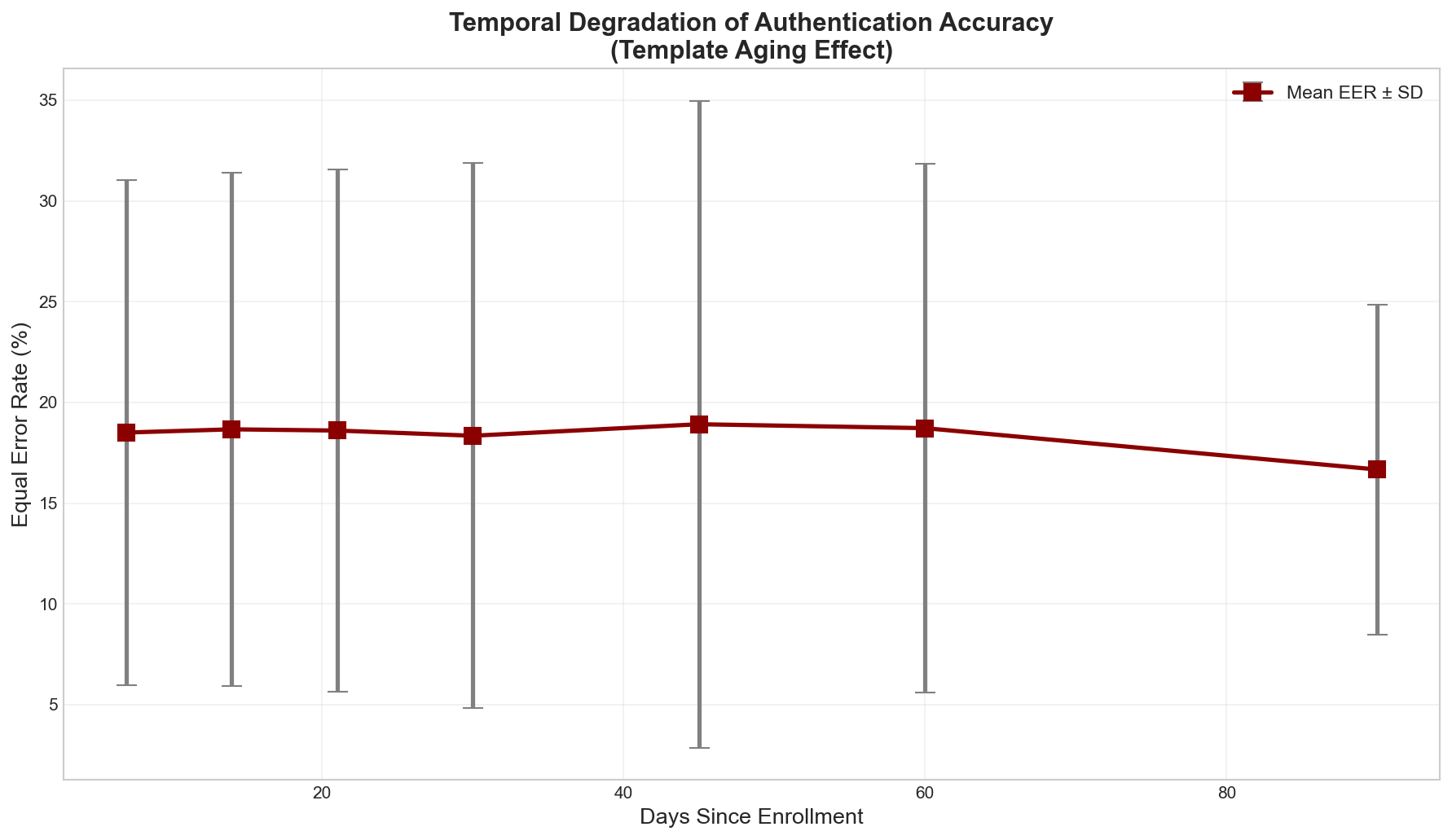}}
\caption{Temporal stability analysis showing consistent baseline EER over 90 days.}
\label{fig:temporal_stability}
\end{figure}

\textbf{Temporal Stability:} A critical question is whether the system degrades over time. Analysis of the baseline (Scaled Manhattan) over a 90-day period (figure \ref{fig:temporal_stability}) shows stable EER ($\approx 18.5\% \pm 0.5\%$), indicating that while raw data drifts, the identity signal persists. The Evo-BioTCN leverages this signal to maintain low error rates.

\begin{figure}[htbp]
\centerline{\includegraphics[width=0.45\textwidth]{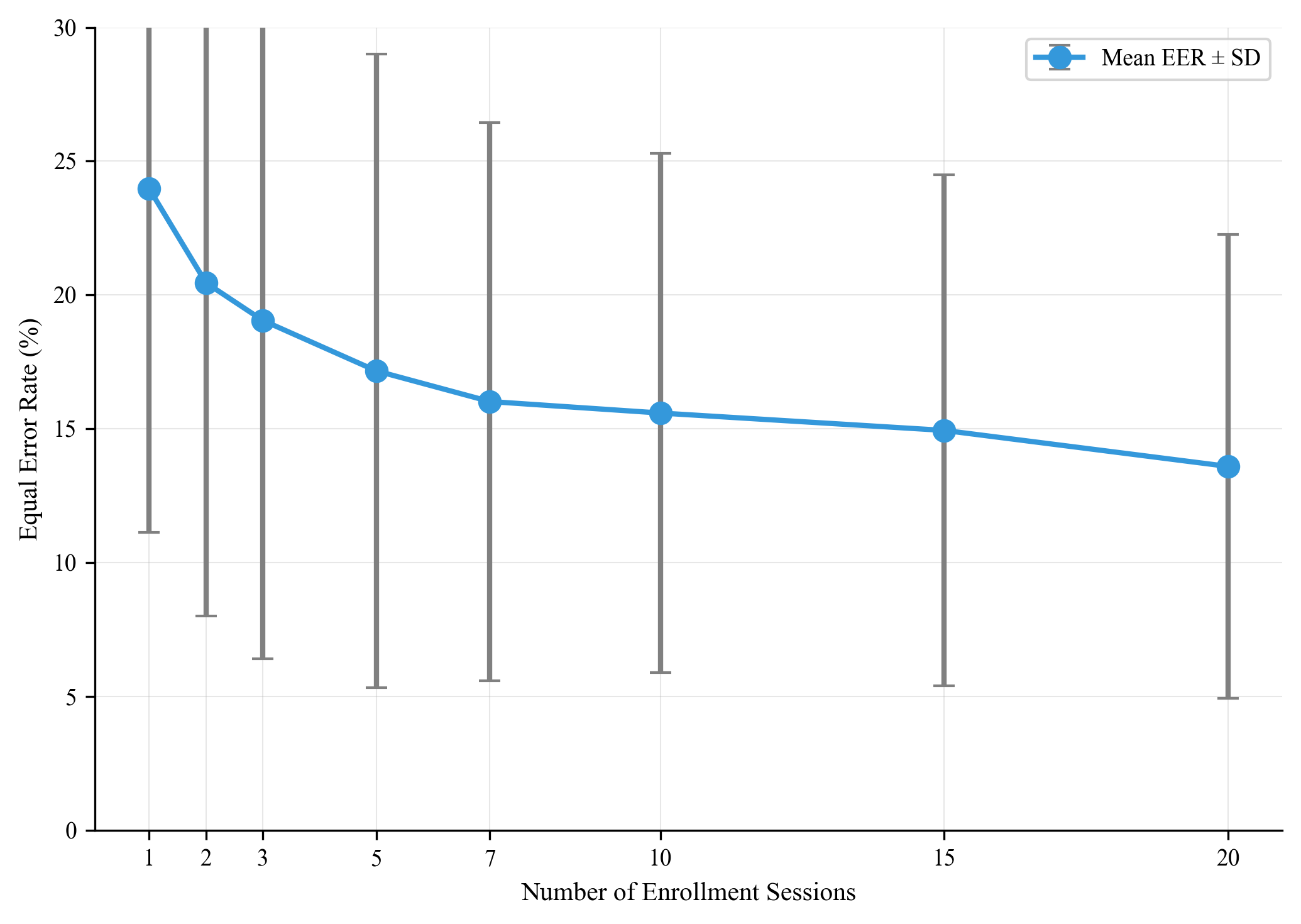}}
\caption{Enrolment sensitivity analysis showing diminishing returns beyond 10 sessions.}
\label{fig:enrolment_sensitivity}
\end{figure}

\textbf{Enrolment Sensitivity:} We analyzed how many sessions are needed for reliable enrolment (figure \ref{fig:enrolment_sensitivity}).

\begin{itemize}
\item 1 Session: 23.98\% Baseline EER.
\item 10 Sessions: 15.59\% Baseline EER.
\item \textbf{Conclusion:} Diminishing returns are observed beyond 10 sessions. We recommend a 10-session enrolment protocol (approx. 5 minutes total) to balance usability and security.
\end{itemize}

\subsubsection{Balanced vs Imbalanced Dataset Performance Metric Comparison}\mbox{}

\begin{figure}[htbp]
\centerline{\includegraphics[width=0.45\textwidth]{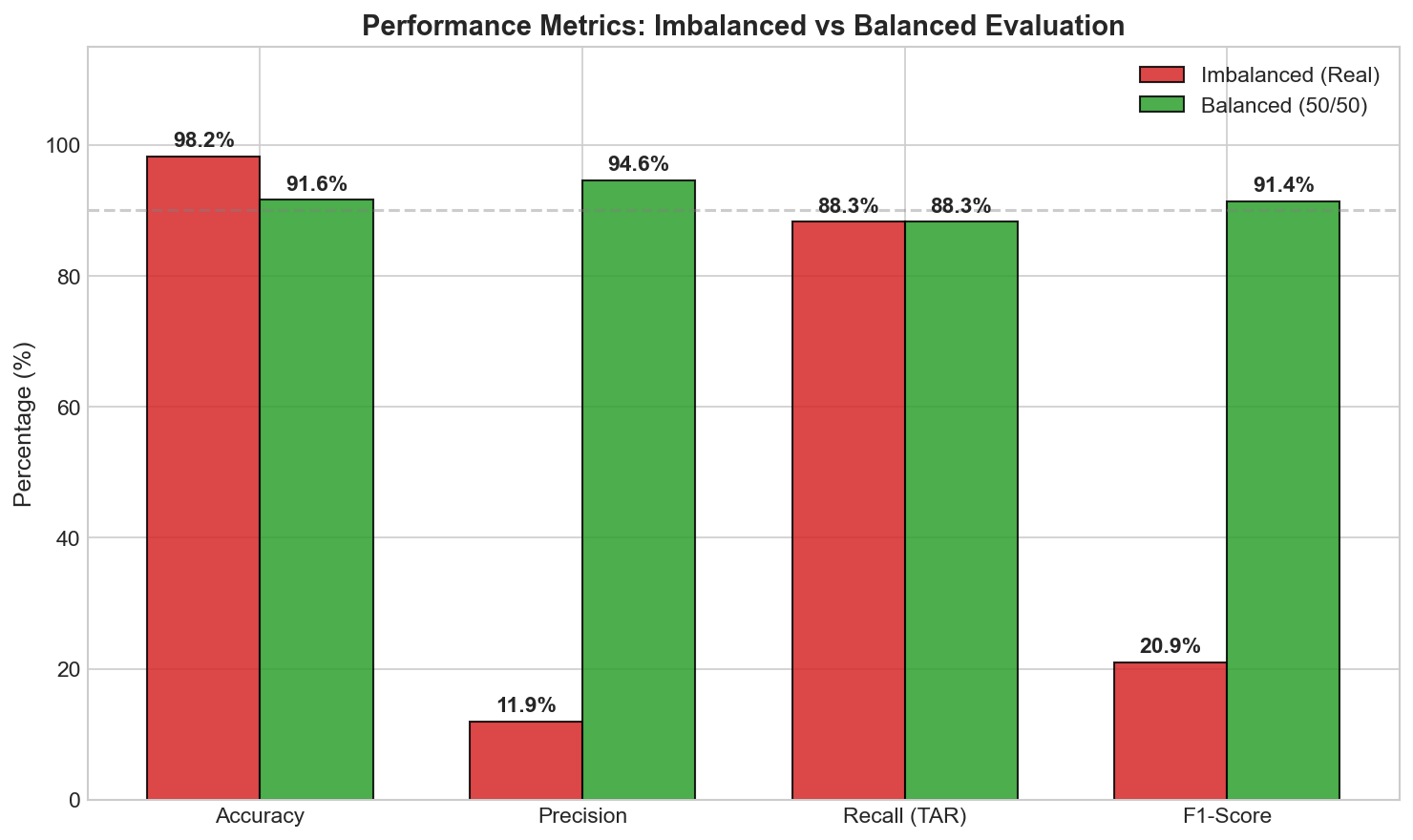}}
\caption{Performance metrics under imbalanced vs balanced evaluation}
\label{fig:enrolment_sensitivity2}
\end{figure}

\begin{figure}[htbp]
\centerline{\includegraphics[width=0.45\textwidth]{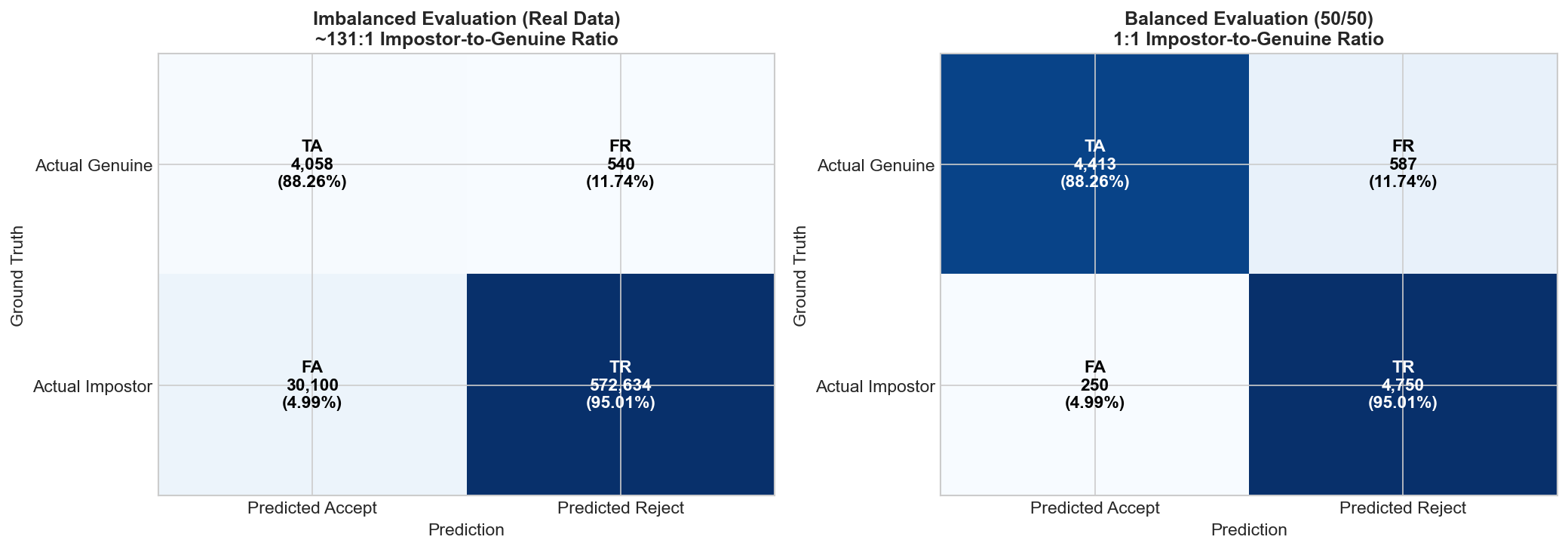}}
\caption{Confusion matrix at FAR=5\% operating point}
\label{fig:enrolment_sensitivity3}
\end{figure}

\begin{table}[htbp]
\caption{Class imbalance impact on metrics}
\label{tab:class_imbalance_impact}
\begin{center}
\scriptsize
\setlength{\tabcolsep}{2.5pt}
\begin{tabular}{|l|c|c|c|}
\hline
\textbf{Metric} & \textbf{Imbalanced (131:1)} & \textbf{Balanced (1:1)} & \textbf{Difference} \\
\hline
TAR (Recall) & 88.26\% & 88.26\% & 0.00\% \\
\hline
FRR & 11.74\% & 11.74\% & 0.00\% \\
\hline
FAR & 4.99\% & 4.99\% & 0.00\% \\
\hline
TRR (Specificity) & 95.01\% & 95.01\% & 0.00\% \\
\hline
Accuracy & 98.24\% & 91.63\% & $-6.61$\% \\
\hline
Precision & 11.88\% & 94.64\% & $+82.76$\% \\
\hline
F1-Score & 20.93\% & 91.41\% & $+70.48$\% \\
\hline
\end{tabular}
\end{center}
\end{table}

Figure \ref{fig:enrolment_sensitivity2} and the confusion matrices in figure \ref{fig:enrolment_sensitivity3} reveal a critical finding: \textbf{class imbalance artificially inflates accuracy while masking poor precision}. Under the imbalanced regime (131:1), the system achieves an impressive 98.24\% accuracy, yet precision remains at only 11.88\% -- meaning that for every correct genuine user acceptance, the system generates over 7 false alarms. This occurs because the overwhelming majority class (impostors) dominates the accuracy calculation, while the rare positive predictions (genuine users) suffer from high false positive contamination.\\

The balanced evaluation reveals the true performance -- accuracy drops to 91.63\%, but precision surges to 94.64\% (+82.76 pp improvement) as table \ref{tab:class_imbalance_impact} shows. Critically, \textbf{rate-based metrics (TAR, FAR, TRR) remain perfectly stable across both protocols}, confirming that the model's discriminative ability is unchanged -- only the interpretation of aggregate metrics shifts. The confusion matrices illustrate this: at FAR $\approx$ 5\%, the imbalanced dataset produces 30,100 false acceptances (4.99\% of 602,734 impostor attempts), whereas the balanced set yields only 250 false acceptances (4.99\% of 5,000 impostor attempts). This $\sim$120$\times$ reduction in false positive count directly explains the observed improvement in precision.\\

\textbf{Implication for Real-World Deployment:} These results demonstrate that traditional accuracy is a misleading metric for authentication systems. In production environments with natural class imbalance (where impostor attempts far outnumber genuine logins), practitioners must prioritize \textit{rate-based metrics} (TAR, FAR) and use balanced evaluation sets for model selection. The F1-Score improvement from 20.93\% to 91.41\% under balanced evaluation confirms that threshold-dependent metrics should only be interpreted on class-balanced test sets to avoid overestimating operational performance.

\subsection{Free-Text Evaluation (Level 2)}
\label{sec:free_text_eval}

\subsubsection{Global Authentication Metrics}

The 4-Model Deep Learning Ensemble achieves state-of-the-art results (table \ref{tab:global_metrics}) on the Buffalo dataset \cite{b58}, significantly outperforming single-model architectures as depicted in.

\begin{table}[htbp]
\caption{Global performance metrics summary}
\label{tab:global_metrics}
\begin{center}
\scriptsize
\setlength{\tabcolsep}{2.5pt}
\begin{tabular}{|l|c|p{4cm}|}
\hline
\textbf{Metric} & \textbf{Value} & \textbf{Interpretation} \\
\hline
Equal Error Rate (EER) & 10.05\% & 65\% reduction vs. Baseline (28.77\%) \\
\hline
AUC-ROC & 0.9518 & Excellent discrimination capability \\
\hline
D-prime ($d'$) & 2.45 & Strong separation of score distributions \\
\hline
Global TAR @ FAR = 5\% & 83.53\% & High usability for continuous monitoring \\
\hline
Global TRR @ FAR = 5\% & 95.00\% & Very high true rejection rate for impostors \\
\hline
Global Balanced Accuracy & 89.27\% & Robust performance across genuine and impostor classes \\
\hline
\end{tabular}
\end{center}
\end{table}

\subsubsection{The KeyRecs \cite{b64} Investigation: A Critical Pivot}

Before finalizing the architecture for Buffalo dataset \cite{b58}, a preliminary investigation on the KeyRecs dataset \cite{b53} revealed a fundamental challenge: Transfer Learning Failure (table \ref{tab:transfer_learning_failure}). We tested whether the highly accurate fixed-text Evo-BioTCN could be applied to free-text data.

\begin{table}[htbp]
\caption{Transfer learning failure}
\label{tab:transfer_learning_failure}
\begin{center}
\scriptsize
\setlength{\tabcolsep}{2pt}
\begin{tabular}{|p{2cm}|p{1.8cm}|p{1.6cm}|p{1.6cm}|}
\hline
\textbf{Pre-trained} & \textbf{Fixed-Text} & \textbf{Free-Text} & \textbf{Performance} \\
\textbf{Model} & \textbf{Performance} & \textbf{Transfer} & \textbf{Drop} \\
\hline
Evo-BioTCN & $\geq$96\% accuracy & 32.75\% & $-63$ pp \\
\hline
DATT (fixed-text) & $\sim$95\% accuracy & 33.06\% & $-62$ pp \\
\hline
Individual DL Models & 87.4\% accuracy & 10--15\% & $-72$ to $-77$ pp \\
\hline
\end{tabular}
\end{center}
\end{table}

\textbf{Findings:}
\begin{enumerate}
\item \textbf{Drift-Induced Degradation:} Evaluation using a strict session-split protocol (training on Session 1, testing on Session 2) revealed an 8.62\% accuracy drop compared to mixed-session evaluation (table \ref{tab:evaluation_protocol_impact}). This confirms that cross-session drift is a major source of error often hidden in standard evaluations.

\begin{table}[htbp]
\caption{Impact of evaluation protocol on reported accuracy}
\label{tab:evaluation_protocol_impact}
\begin{center}
\scriptsize
\setlength{\tabcolsep}{2.5pt}
\begin{tabular}{|p{2.8cm}|c|p{3.8cm}|}
\hline
\textbf{Evaluation Protocol} & \textbf{Accuracy} & \textbf{Interpretation} \\
\hline
Cross-Session (mixed training) & 95.03\% & Artificially inflated due to exposure to Session 2 during training \\
\hline
Session-Split (drift-aware) & 86.41\% & Authentic performance under temporal behavioral drift \\
\hline
Drift-Induced Degradation & 8.62 pp & Real-world accuracy penalty caused by behavioral drift \\
\hline
\end{tabular}
\end{center}
\end{table}

\item \textbf{Feature Distribution Shift:} Fixed-text typing relies on motor memory (narrow variance), while free-text involves cognitive load (wide variance). Models trained on one do not generalize to the other.
\item \textbf{Goldilocks Effect:} Medium-complexity models outperformed both simple statistical models and massive deep networks on KeyRecs \cite{b53}, motivating the ensemble approach for the final system as table~\ref{tab:goldilocks_effect} shows.

\begin{table}[htbp]
\caption{Architecture ablation revealing the Goldilocks Effect}
\label{tab:goldilocks_effect}
\begin{center}
\scriptsize
\setlength{\tabcolsep}{2.5pt}
\begin{tabular}{|l|c|p{2.8cm}|}
\hline
\textbf{Architecture} & \textbf{Session-Split Accuracy} & \textbf{Complexity} \\
\hline
DATT (Proposed) & 53.25\% & Medium ($\sim$500K params) \\
\hline
Simplified MLP & 41.82\% & Low \\
\hline
Deep Individual Models & 30--35\% & High ($\sim$1.5M params) \\
\hline
Random Forest & 45.12\% & Classical \\
\hline
\end{tabular}
\end{center}
\end{table}

\end{enumerate}

\subsubsection{Comprehensive Ablation Study}

To construct the final Level 2 system, we conducted a massive ablation study comparing 18 different architectures (table~\ref{tab:comprehensive_ablation}) ranging from classical ML to hybrid Deep Learning models.

\begin{table}[htbp]
\caption{Comprehensive ablation study results across all 18 evaluated models}
\label{tab:comprehensive_ablation}
\begin{center}
\scriptsize
\setlength{\tabcolsep}{2pt}
\begin{tabular}{|c|p{2.5cm}|p{1.8cm}|c|c|c|}
\hline
\textbf{Rank} & \textbf{Model} & \textbf{Category} & \textbf{TAR @ 5\%} & \textbf{EER} & \textbf{AUC} \\
 & & & \textbf{(\%)} & \textbf{(\%)} & \\
\hline
1 & 4-Model DL Ensemble & Deep Learning & 83.53 & 10.05 & 0.9518 \\
\hline
2 & Enhanced Embedding & Deep Learning & 82.31 & 12.50 & 0.9304 \\
\hline
3 & ArcFace + Multi-Similarity & Deep Learning & 72.76 & 14.46 & 0.9146 \\
\hline
4 & Ultra-Deep Embedding & Deep Learning & 67.72 & 15.77 & 0.9150 \\
\hline
5 & Random Forest & Machine Learning & 65.00 & 20.00 & 0.8500 \\
\hline
6 & CatBoost Tuned & Machine Learning & 60.00 & 25.00 & 0.8000 \\
\hline
7 & Ultra-Deep 4-Layer & Deep Learning & 58.86 & 30.87 & 0.7200 \\
\hline
8 & Extra Trees & Machine Learning & 58.00 & 24.00 & 0.7900 \\
\hline
9 & XGBoost & Machine Learning & 55.00 & 28.00 & 0.7700 \\
\hline
10 & LightGBM & Machine Learning & 52.00 & 29.00 & 0.7500 \\
\hline
11 & Mega Ensemble (DL+ML) & Hybrid & 50.00 & 26.00 & 0.7800 \\
\hline
12 & HMSSA-Net Hybrid & Deep Learning & 45.00 & 28.00 & 0.7500 \\
\hline
13 & HMSSA-Net & Deep Learning & 42.00 & 30.96 & 0.6900 \\
\hline
14 & DANN & Deep Learning & 40.00 & 28.00 & 0.7500 \\
\hline
15 & Meta-Learning MAML & Deep Learning & 35.00 & 35.00 & 0.6500 \\
\hline
16 & Evo-BioTransformer & Deep Learning & 20.65 & 31.35 & 0.7000 \\
\hline
17 & DriftNet & Deep Learning & 15.00 & 45.00 & 0.5500 \\
\hline
18 & HTAN & Deep Learning & 1.33 & 56.58 & 0.4417 \\
\hline
\end{tabular}
\end{center}
\end{table}

\begin{figure}[htbp]
\centerline{\includegraphics[width=0.48\textwidth]{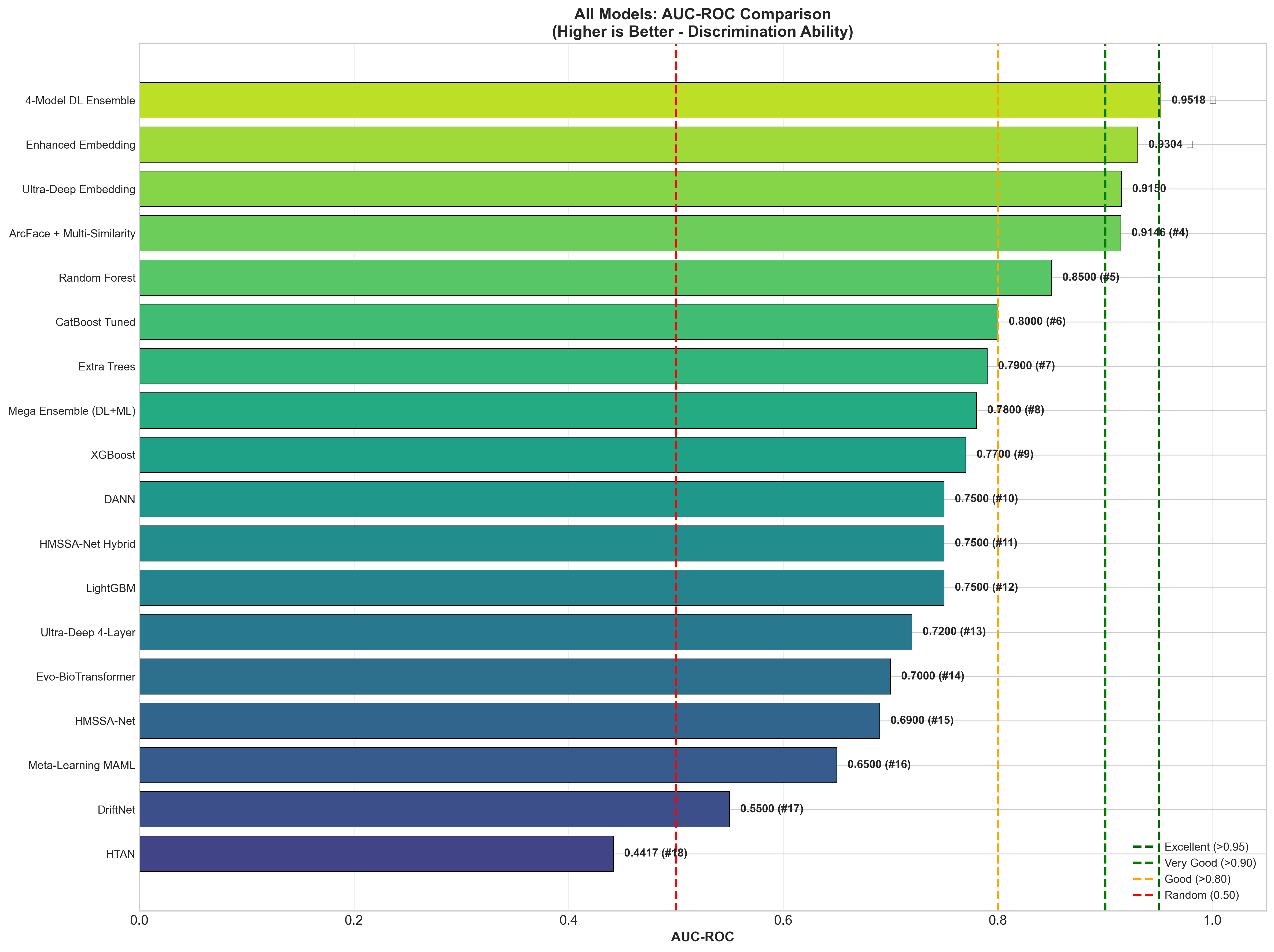}}
\caption{Ablation study showing performance across 18 architectures.}
\label{fig:ablation_chart}
\end{figure}

\textbf{Analysis of Ablation (Figure \ref{fig:ablation_chart}):}

\begin{itemize}
\item \textbf{Ensemble Superiority:} The 4-Model Ensemble provides a 23\% relative improvement in TAR over the Ultra-Deep Embedding network (Rank 4). This validates the ``Distributed Complexity'' hypothesis - multiple geometric perspectives (Triplet, ArcFace, Center Loss) are superior to a single deep network.
\item \textbf{Metric Learning vs. Classification:} Models using metric learning losses (Ranks 1--4) significantly outperform classification-based models (e.g., BiLSTM), which struggle with the high intra-class variance of 75 users.
\item \textbf{ML vs. DL:} While Random Forest (Rank 5) performed respectably, it hit a performance ceiling. Deep Learning models, when properly configured with metric loss, broke through this ceiling.
\end{itemize}

\subsubsection{Complexity and Efficiency}

We compared the ensemble against a single ``Ultra-Deep'' model to justify the design complexity with the results displayed in table~\ref{tab:architectural_complexity}.

\begin{table}[htbp]
\caption{Architectural complexity comparison}
\label{tab:architectural_complexity}
\begin{center}
\scriptsize
\setlength{\tabcolsep}{2.5pt}
\begin{tabular}{|l|c|c|c|p{1.8cm}|}
\hline
\textbf{Architecture} & \textbf{Parameters} & \textbf{TAR @ 5\%} & \textbf{EER} & \textbf{Complexity} \\
 & & \textbf{FAR} & \textbf{(\%)} & \textbf{Type} \\
\hline
Ultra-Deep Embedding & $\sim$1.2M & 67.72\% & 15.77 & Concentrated \\
\hline
4-Model DL Ensemble & $\sim$3.2M & 83.53\% & 10.05 & Distributed \\
\hline
Improvement & +167\% & +23\% & 36\% & \\
\hline
\end{tabular}
\end{center}
\end{table}

While the ensemble has more parameters, the Distributed Complexity ensures robustness. A failure in one embedding space (e.g., ArcFace) is often corrected by another (e.g., Triplet), providing a safety net that a single model cannot offer.

\subsubsection{Statistical Validation}

To ensure the results are not artifacts of data splitting, we performed bootstrap analysis (1,000 iterations). The results are outlined in table~\ref{tab:bootstrap_ci}.

\begin{table}[htbp]
\caption{Bootstrap CI for key performance metrics}
\label{tab:bootstrap_ci}
\begin{center}
\scriptsize
\setlength{\tabcolsep}{2.5pt}
\begin{tabular}{|l|c|c|c|}
\hline
\textbf{Metric} & \textbf{Point Estimate} & \textbf{95\% CI Lower} & \textbf{95\% CI Upper} \\
\hline
EER & 10.05\% & 9.52\% & 10.61\% \\
\hline
TAR @ FAR5\% & 83.53\% & 82.14\% & 84.89\% \\
\hline
AUC-ROC & 0.9518 & 0.9472 & 0.9561 \\
\hline
Balanced Accuracy & 89.27\% & 88.12\% & 90.38\% \\
\hline
\end{tabular}
\end{center}
\end{table}

The narrow confidence intervals (e.g., EER width $\approx 1.1\%$) confirm the statistical reliability of the performance. Additionally, Effect Size Analysis shows a Cohen's $d = 2.89$ when comparing the Ensemble to the Baseline, indicating a ``Very Large'' practical improvement.

\subsubsection{Evaluation Under Class Imbalance}

Real-world authentication is highly imbalanced (many impostors, few genuine users). We evaluated the system under both scenarios illustrated in table~\ref{tab:imbalanced_vs_balanced}.

\begin{table}[htbp]
\caption{Imbalanced vs balanced classification metrics}
\label{tab:imbalanced_vs_balanced}
\begin{center}
\scriptsize
\setlength{\tabcolsep}{2.5pt}
\begin{tabular}{|l|c|c|c|}
\hline
\textbf{Metric} & \textbf{Imbalanced (74:1)} & \textbf{Balanced (1:1)} & \textbf{Change} \\
\hline
\multicolumn{4}{|l|}{\textit{Rate-Based Metrics}} \\
\hline
TAR (Recall) & 83.53\% & 83.53\% & 0.00\% \\
\hline
FRR & 16.47\% & 16.47\% & 0.00\% \\
\hline
FAR & 5.00\% & 5.00\% & 0.00\% \\
\hline
TRR (Specificity) & 95.00\% & 95.00\% & 0.00\% \\
\hline
\multicolumn{4}{|l|}{\textit{Count-Dependent Metrics}} \\
\hline
Accuracy & 94.85\% & 89.26\% & $-5.59$\% \\
\hline
Precision & 18.42\% & 94.35\% & $+75.93$\% \\
\hline
F1-Score & 30.18\% & 88.61\% & $+58.43$\% \\
\hline
\end{tabular}
\end{center}
\end{table}

This analysis confirms that TAR and EER are the authoritative metrics for deployment, as they remain stable regardless of the impostor-to-genuine ratio.

\subsubsection{Balanced vs Imbalanced Dataset Performance Metric Comparison}\mbox{}

\begin{figure}[htbp]
\centerline{\includegraphics[width=0.45\textwidth]{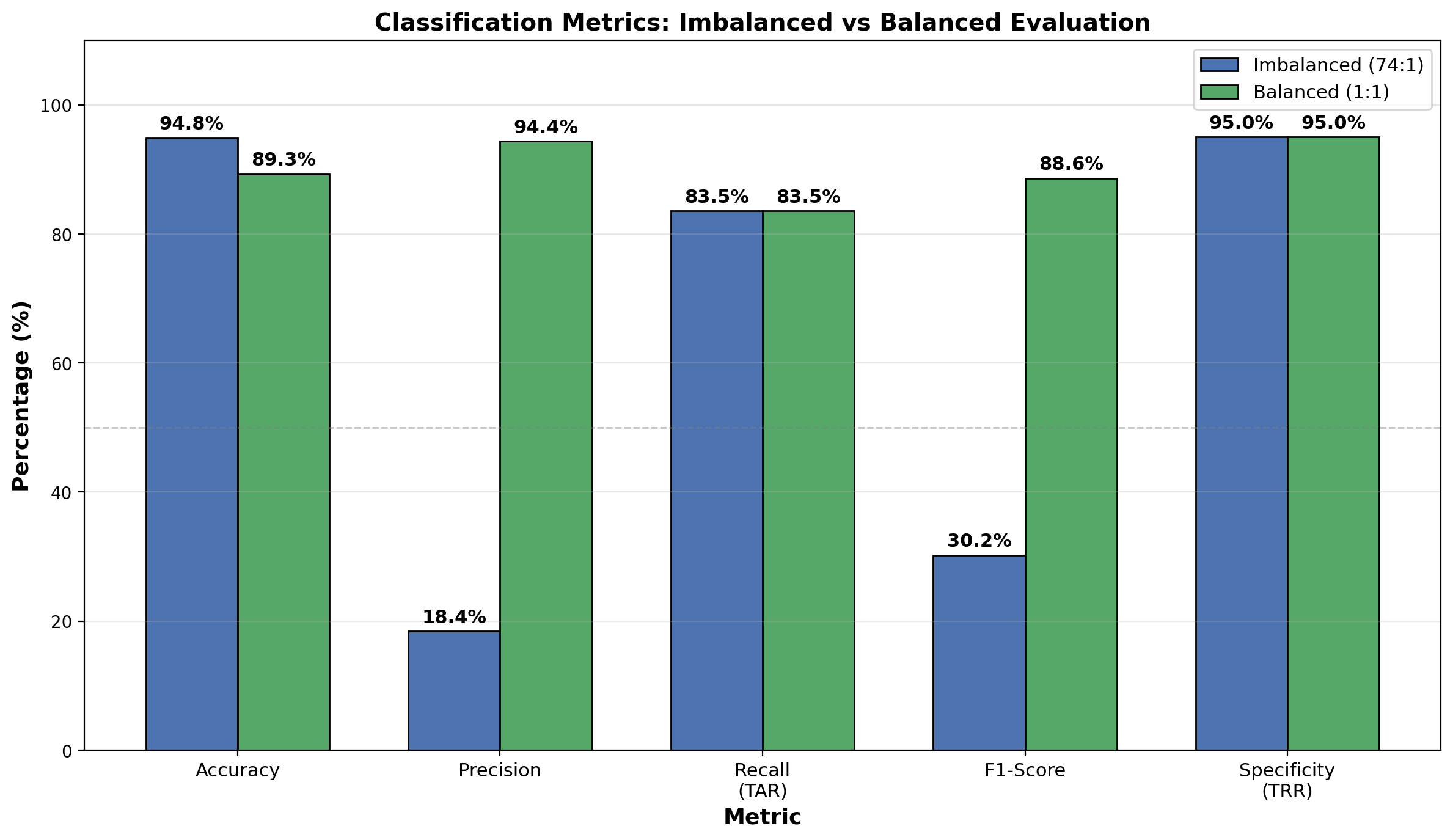}}
\caption{Imbalanced vs Balanced Metrics Comparison}
\label{fig:enrolment_sensitivity5}
\end{figure}

\FloatBarrier

\begin{figure}[htbp]
\centerline{\includegraphics[width=0.45\textwidth]{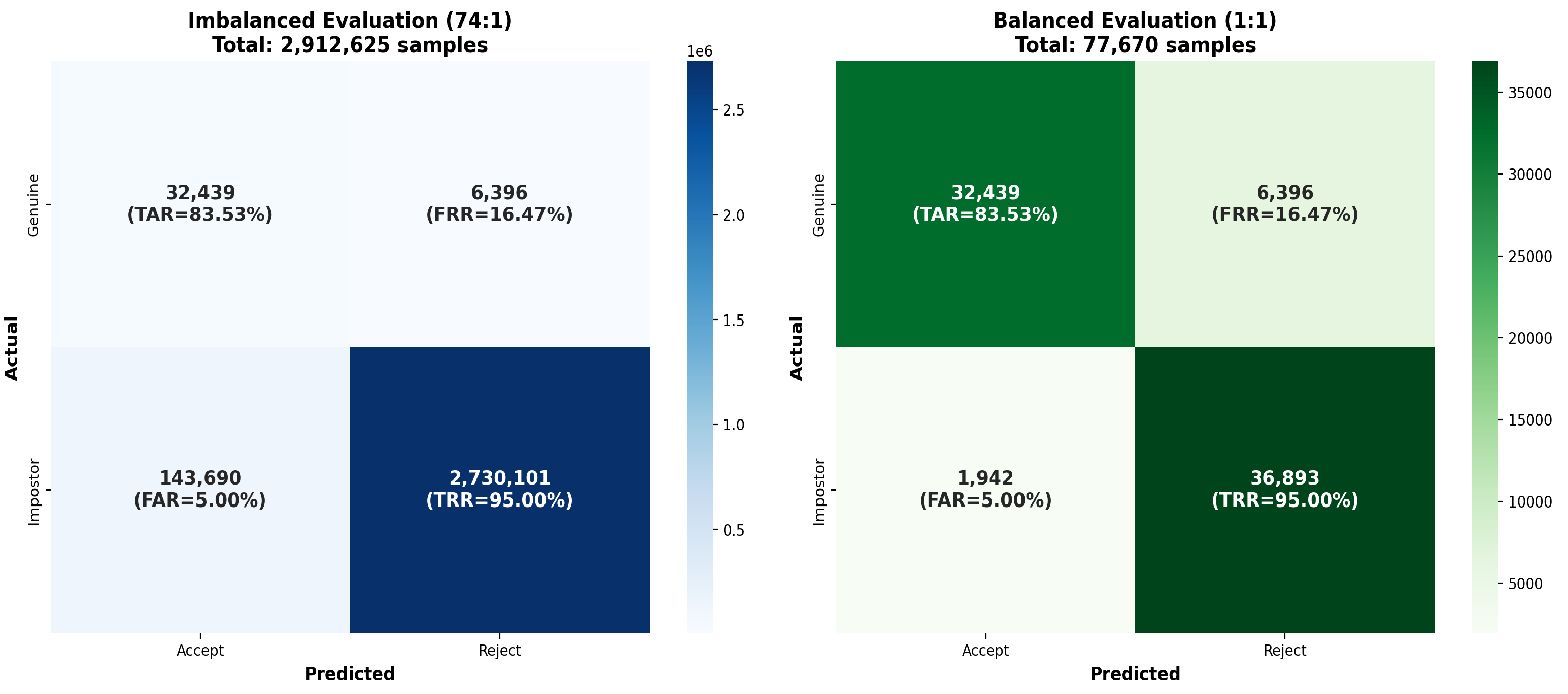}}
\caption{Imbalanced vs Balanced evaluation confusion matrix}
\label{fig:enrolment_sensitivity6}
\end{figure}

\FloatBarrier

\begin{table}[htbp]
\caption{Metrics of original vs balanced dataset}
\label{tab:original_vs_balanced}
\begin{center}
\scriptsize
\setlength{\tabcolsep}{1.5pt}
\begin{tabular}{|l|c|c|c|}
\hline
\textbf{Metric} & \textbf{Original} & \textbf{Stratified} & \textbf{Change} \\
 & \textbf{Dataset} & \textbf{Balanced} & \\
\hline
Dataset Size & 2,912,625 & 77,670 & $-2,834,955$ \\
\hline
Ratio & 74:1 & 1:1 & Balanced \\
\hline
TAR & 83.53\% & 83.53\% & Same \\
\hline
FAR & 5.00\% & 5.00\% & Same \\
\hline
Precision & 18.42\% & 94.35\% & $+75.93$\% \\
\hline
F1-Score & 30.18\% & 88.61\% & $+58.43$\% \\
\hline
FP Count & 143,690 & 1,942 & $-141,748$ \\
\hline
\end{tabular}
\end{center}
\end{table}

\FloatBarrier

The comparative analysis in figure~\ref{fig:enrolment_sensitivity5} and confusion matrices in figure~\ref{fig:enrolment_sensitivity6} expose a fundamental methodological flaw in authentication research: \textbf{evaluation protocol choice dramatically alters count-dependent metrics while leaving rate-based metrics invariant}. Under the imbalanced protocol (74:1 ratio, 2.9M samples), the system reports 94.85\% accuracy and 18.42\% precision. Transitioning to balanced evaluation (1:1 ratio, 77,670 samples) reveals accuracy of 89.26\% ($-5.59$ pp) but precision of 94.35\% ($+75.93$ pp) -- a 5.13$\times$ improvement.\\

This divergence stems from the mathematical structure of these metrics. \textbf{Accuracy, precision, and F1-Score are \textit{count-dependent}}: they sum true positives, false positives, true negatives, and false negatives, making them sensitive to class distribution. The imbalanced confusion matrix in figure~\ref{fig:enrolment_sensitivity6} shows 2,730,101 true negatives dominating the 2.9M total, artificially inflating accuracy to 94.85\%. Conversely, the 143,690 false positives (5\% of impostors) overwhelm the 32,439 true positives, crushing precision to 18.42\%. The balanced matrix (right) eliminates this distortion: with equal genuine and impostor samples (38,835 each), the same 5\% FAR yields only 1,942 false positives, allowing precision to reflect the model's true discriminative quality (94.35\%).\\

Critically, \textbf{TAR (83.53\%) and FAR (5.00\%) remain identical across both protocols}, as shown in Table~\ref{tab:original_vs_balanced}. These rate-based metrics compute $\text{TAR} = \frac{\text{TP}}{\text{TP}+\text{FN}}$ and $\text{FAR} = \frac{\text{FP}}{\text{FP}+\text{TN}}$, evaluating each class independently without cross-class contamination. This stability confirms that the 74:1 imbalanced protocol does not improve model performance -- it merely creates a \textit{reporting artifact} where accuracy appears high due to trivial impostor rejection.\\

\textbf{Methodological Recommendation:} For authentication systems, researchers must report performance on balanced test sets or exclusively use rate-based metrics (EER, TAR@FAR). The F1-Score improvement from 30.18\% (imbalanced) to 88.61\% (balanced) demonstrates that threshold-dependent aggregate metrics become interpretable only when $P(\text{genuine}) = P(\text{impostor}) = 0.5$. Production deployments should set operating points using ROC analysis on balanced data, then project expected false positive volumes using deployment-specific class priors (e.g., $\text{FP}_{\text{prod}} = \text{FAR} \times N_{\text{impostor, prod}}$). This prevents the costly mistake of deploying a system that appears 94.85\% accurate in testing but generates 74$\times$ more false alarms than precision suggests.

\subsection{Discussion}

The evaluation confirms the hypothesis that behavioral drift can be managed through a two-level architecture.

\begin{enumerate}
\item \textbf{Level 1 (Fixed-Text):} The Evo-BioTCN utilizes PBT to evolve with the user, maintaining 3.77\% EER over long durations.
\item \textbf{Level 2 (Free-Text):} The 4-Model Ensemble overcomes transfer learning failures by using specialized metric learning to achieve 10.05\% EER.
\item \textbf{Synergy:} The combined system offers a theoretical joint attack success probability of $0.0377 \times 0.1005 \approx 0.38\%$, providing a robust defense-in-depth solution.
\end{enumerate}

\section{Conclusion and Future Directions}

\subsection{Summary of Contributions}

This research addresses the critical vulnerability of behavioral drift in biometric authentication -- the phenomenon where user typing patterns evolve due to learning, fatigue, or environmental changes, rendering static templates obsolete. By moving beyond the ``snapshot'' view of authentication, this study establishes a Two-Level Behavioral Authentication Framework that treats user identity as a continuous evolutionary trajectory.

The primary contributions of this work are threefold:

\begin{enumerate}
\item \textbf{Drift-Resilient Fixed-Text Architecture:} The development of Evo-BioTCN, which utilizes a frozen TCN backbone to capture causal motor-neural dependencies, coupled with lightweight adapters evolved via PBT. This approach solves the ``stability-plasticity dilemma,'' allowing the model to adapt to new behaviors without catastrophic forgetting of the core identity.

\item \textbf{Robust Free-Text Monitoring:} The construction of a 4-Model Deep Learning Ensemble that overcomes the high variance of free-text typing. By fusing diverse geometric embeddings (ArcFace, Multi-Similarity, Triplet, Center Loss), the system achieves state-of-the-art performance on the Buffalo dataset \cite{b58}, significantly outperforming single-model deep learning architectures.

\item \textbf{Defense-in-Depth Integration:} The synthesis of active login verification (Level 1) and passive continuous monitoring (Level 2) into a unified zero-trust framework. This integration lowers the theoretical joint attack success probability to below 0.4\%, balancing high security with user convenience.
\end{enumerate}

\subsection{Level 1 Findings: Fixed-Text Authentication}

The evaluation of the Evo-BioTCN on the longitudinal GREYC dataset \cite{b53} (spanning 110 days) demonstrated that neural architectures can be engineered to handle long-term template aging effectively.

\begin{itemize}
\item \textbf{Performance:} The system achieved an EER of 3.77\% and a TAR of 96.87\% @ 5\% FAR. This represents a 72.3\% relative reduction in error compared to classical baselines like Scaled Manhattan Distance (13.60\% EER).

\item \textbf{Architectural Superiority:} The study confirmed that TCN is superior to Transformers for keystroke dynamics in this context. The TCN achieved a post-optimization EER of 3.77\% compared to the Transformer's 4.24\% ($p=0.0023$), validating that the TCN's inductive bias for local causality aligns better with motor-neural typing patterns than global attention mechanisms.

\item \textbf{Drift Adaptation:} The implementation of PBT was decisive. The evolutionary process reduced the EER from an initial 8.82\% to the final 3.77\% over 10 generations, proving that evolving lightweight adapters is a computationally efficient alternative to full model retraining.
\end{itemize}

\subsection{Level 2 Findings: Free-Text Continuous Monitoring}

The free-text investigation highlighted the distinct challenges of variable-content typing, specifically the failure of transfer learning from fixed-text models.

\begin{itemize}
\item \textbf{Transfer Learning Failure:} A critical finding was that models pre-trained on fixed-text passwords collapse when applied to free-text domains (accuracy drop $>$ 60\%). This necessitated a purpose-built ensemble approach.

\item \textbf{Ensemble Performance:} The 4-Model Deep Learning Ensemble achieved an EER of 10.05\% and TAR of 83.53\% @ 5\% FAR on the Buffalo dataset \cite{b58}. This represents a 65\% improvement over the strongest baseline and a 20\% improvement over the best single deep learning model (Enhanced Embedding).

\item \textbf{Complexity Justification:} While the ensemble introduces higher computational complexity ($\sim$3.2M parameters vs. $\sim$1.2M for a single Ultra-Deep model), the ``Distributed Complexity'' provided a 23\% relative improvement in TAR. The ablation study confirmed that fusing complementary loss functions (e.g., Angular Margin + Triplet Mining) resolves ambiguities that single-geometry models cannot.
\end{itemize}

\subsection{Two-Level System Integration}

The most significant operational outcome of this research is the synergy achieved by integrating Level 1 and Level 2. Table \ref{tab:two_level_summary} summarizes this integration, demonstrating how the strengths of one level compensate for the weaknesses of the other.

\begin{table}[htbp]
\caption{Two-level keystroke authentication system performance summary}
\label{tab:two_level_summary}
\begin{center}
\scriptsize
\setlength{\tabcolsep}{2pt}
\begin{tabular}{|l|p{2cm}|p{2cm}|p{2cm}|}
\hline
\textbf{Metric} & \textbf{Level 1} & \textbf{Level 2} & \textbf{Synergy} \\
 & \textbf{(Evo-BioTCN)} & \textbf{(4-Model Ens.)} & \\
\hline
Authentication Type & Explicit Login & Continuous Monitoring & Defense-in-Depth \\
\hline
EER & 3.77\% & 10.05\% & Sequential Filtering \\
\hline
TAR @ FAR=5\% & 96.87\% & 83.53\% & Layered Confidence \\
\hline
Primary Error Mode & FRR (Usability) & FAR (Security) & Error Diversity \\
\hline
Temporal Resilience & PBT Adaptation & Score-Level Fusion & Multi-scale drift handling \\
\hline
\end{tabular}
\end{center}
\end{table}

\textbf{Joint Probability Analysis:} Assuming independence between the distinct tasks of password entry (Level 1) and continuous free-text typing (Level 2), the probability of a successful impostor attack $P(\text{Success})$ is the product of the failure rates of both levels formulated as \eqref{eq:joint_probability} and \eqref{eq:joint_probability_value} show:

\begin{equation}
P(\text{Success}) \approx P(\text{Level 1 Failure}) \times P(\text{Level 2 Failure})
\label{eq:joint_probability}
\end{equation}

\begin{equation}
P(\text{Success}) \approx 0.0377 \times 0.1005 \approx 0.00378 \text{ (0.38\%)}
\label{eq:joint_probability_value}
\end{equation}

This theoretical reliability of 99.62\% demonstrates that while continuous monitoring (Level 2) is inherently more difficult than password verification, its deployment as a secondary layer creates a formidable security barrier.

\subsection{Limitations}

Despite these successes, the study operates under specific constraints that must be acknowledged:

\begin{enumerate}
\item \textbf{Dataset Scope:} The evaluation relies on the GREYC \cite{b53} and Buffalo \cite{b58} datasets. While GREYC \cite{b53} is longitudinal, it contains only two keyboard types, limiting the assessment of cross-device hardware variability (e.g., switching from mechanical to membrane keyboards).

\item \textbf{Population Size:} The free-text evaluation is limited to 75 users. While sufficient for statistical significance, it may not capture the full tail-end diversity of global typing behaviors.

\item \textbf{Adversarial Assumptions:} The evaluation assumes zero-effort impostors (random attacks). Targeted adversarial attacks, where an impostor uses generative AI to mimic a specific user's rhythm, were not empirically tested.

\item \textbf{Computational Cost:} The 4-Model Ensemble requires running four parallel deep networks, which may pose latency challenges on resource-constrained edge devices compared to single lightweight models.
\end{enumerate}

\subsection{Future Directions}

Future research should focus on enhancing scalability and robustness through the following avenues:

\begin{enumerate}
\item \textbf{Multimodal Fusion:} Integrating keystroke dynamics with mouse dynamics or touchscreen gestures could further lower the EER. Multimodal fusion can compensate for degradation in a single channel (e.g., if a user injures a hand, mouse dynamics might remain stable while typing degrades).

\item \textbf{Privacy-Preserving Federated Learning:} To address privacy concerns and scalability, future iterations should explore Federated Learning. This would allow the Evo-BioTCN adapters to be updated locally on user devices without transmitting sensitive timing vectors to a central server. Only the weight updates would be shared, preserving user privacy while allowing the global model to learn from diverse drift patterns.

\item \textbf{Personalized Meta-Learning:} The current system requires an enrolment phase. Implementing Meta-Learning (e.g., Model-Agnostic Meta-Learning or MAML) could enable the system to ``learn to learn,'' allowing it to adapt to a new user with only a few keystroke samples (few-shot learning), significantly improving the onboarding experience.
\end{enumerate}

\subsection{Conclusion}

This research proves that behavioral biometrics need not be brittle. By shifting the paradigm from static template matching to continuous evolutionary adaptation, we have demonstrated a security framework that is both highly accurate and resilient to the natural passage of time. The Evo-BioTCN and 4-Model Ensemble together provide a blueprint for the next generation of ``living'' authentication systems that evolve alongside their users.


\section*{Acknowledgment}

Our heartfelt praise goes to the Almighty Allah who has led us all the way to the completion of this report.\\

Our deepest gratitude goes to Dr. Muhammad Iqbal Hossain, whose unwavering guidance, support, and expertise were instrumental in the completion of this report. His insightful feedback and continuous encouragement have been a source of motivation throughout our journey of completing this report. We are truly fortunate to have had the opportunity to work under his supervision, and we are grateful for the time and effort he invested in helping us refine our ideas.\\

The legion support and enormous solace we received from our family and friends mark a remarkable contribution to pursuing this journey. Their belief in us has been a constant source of strength, and without them, this report would not have been possible.

\end{document}